 \documentclass[final,5p,times,twocolumn,authoryear]{elsarticle}

\usepackage{amssymb}
\usepackage[dvipsnames]{xcolor}
\usepackage{rotating}
\usepackage{threeparttable}
\usepackage{multirow}
\usepackage{hyperref}
\usepackage{lineno}
\usepackage{graphicx}
\usepackage{xcolor}
\usepackage{pdflscape}
\usepackage[utf8]{inputenc}
\usepackage[T1]{fontenc}    
\usepackage{caption}
\usepackage{subcaption}
\usepackage{url}
\usepackage{amsmath}	
\usepackage{txfonts}
\usepackage{setspace}
\usepackage[normalem]{ulem}
\usepackage{array}
\setcitestyle{aysep={}} 

\usepackage{hyperref}
\hypersetup{breaklinks=true}
\usepackage{url}
\usepackage{breakurl} 

\newcommand{\notes}[1]{{}}
\newcommand{\review}[1]{{#1}}
\newcommand{\colreview}[1]{{#1}}

\newcommand{\photoz}{$z_{phot}$ }

\newcommand{\specz}{$z_{spec}$ }

\journal{Astronomy $\&$ Computing}

\begin{document}

\begin{frontmatter}

\title{Photometric Redshifts Probability Density Estimation from Recurrent Neural Networks in the DECam Local Volume Exploration Survey Data Release 2}

\renewcommand{\thefootnote}{\alph{footnote}}  

\author[CBPF]{G. Teixeira\corref{cor1}}
\ead{gteixeira@cbpf.br}
\author[CBPF,CEFET]{C. R. Bom} 
\author[CBPF]{L. Santana-Silva} 
\author[CBPF]{B.M.O. Fraga} 
\author[CBPF]{P. Darc} 
\author[UChicago]{R. Teixeira} 
\author[STScI,CASDP]{J.~F.~Wu} 
\author[UW]{P.~S.~Ferguson}
\author[IGO]{C.~E.~Martínez-Vázquez}
\author[ICC]{A.~H.~Riley}
\author[Fermilab,KICP,DAA]{A.~Drlica-Wagner}
\author[NSF]{Y.~Choi}
\author[Dartmouth]{B.~Mutlu-Pakdil}
\author[CMU]{A.~B.~Pace}
\author[Surrey]{J.~D.~Sakowska}
\author[CASA]{G.~S.~Stringfellow}

\address[CBPF]{Centro Brasileiro de Pesquisas Físicas, Rua Dr. Xavier Sigaud 150, 22290-180 Rio de Janeiro, RJ, Brazil}
\address[CEFET]{Centro Federal de Educa\c{c}\~{a}o Tecnol\'{o}gica Celso Suckow da Fonseca,  Rodovia M\'{a}rcio Covas, lote J2, quadra J - Itagua\'{i} (Brazil)}
\address[UChicago]{Department of Astronomy and Astrophysics, University of Chicago, 5640 S Ellis Ave, Chicago, IL 60637, USA}
\address[STScI]{Space Telescope Science Institute, 3700 San Martin Drive, Baltimore, MD 21218, USA}
\address[CASDP]{Center for Astrophysical Sciences, Department of Physics \& Astronomy, Johns Hopkins University, Baltimore, MD 21218, USA}
\address[UW]{Department of Physics, University of Wisconsin-Madison, Madison, WI 53706, USA}
\address[IGO]{International Gemini Observatory/NSF NOIRLab, 670 N. A'ohoku Place, Hilo, Hawai'i, 96720, USA}
\address[ICC]{Institute for Computational Cosmology, Department of Physics, Durham University, South Road, Durham DH1 3LE, UK}
\address[Fermilab]{Fermi National Accelerator Laboratory, P.O. Box 500, Batavia, IL 60510, USA}
\address[KICP]{Kavli Institute for Cosmological Physics, University of Chicago, Chicago, IL 60637, USA}
\address[DAA]{Department of Astronomy and Astrophysics, University of Chicago, Chicago, IL 60637, USA}
\address[NSF]{NSF NOIRLab, 950 N. Cherry Ave., Tucson, AZ 85719, USA}
\address[Dartmouth]{Department of Physics and Astronomy, Dartmouth College, Hanover, NH 03755, USA}
\address[CMU]{McWilliams Center for Cosmology \& Astrophysics, Carnegie Mellon University, 5000 Forbes Ave, Pittsburgh, PA 15213, USA}
\address[Surrey]{Department of Physics, University of Surrey, Guildford GU2 7XH, UK}
\address[CASA]{Center for Astrophysics and Space Astronomy, University of Colorado, 389 UCB, Boulder, CO 80309-0389, USA}

\cortext[cor1]{Corresponding author}

\begin{abstract}
    Photometric wide-field surveys are imaging the sky in unprecedented detail. These surveys face a significant challenge in efficiently estimating galactic photometric redshifts while accurately quantifying associated uncertainties. 
    In this work, we address this challenge by exploring the estimation of Probability Density Functions (PDFs) for the photometric redshifts of galaxies across a vast area of 17,000 square degrees, encompassing objects with a median 5$\sigma$ point-source depth of $g$ = 24.3, $r$ = 23.9, $i$ = 23.5, and $z$ = 22.8 mag.
    
   Our approach uses deep learning, specifically integrating a Recurrent Neural Network architecture with a Mixture Density Network, to leverage magnitudes and colors as input features for constructing photometric redshift PDFs across the whole DECam Local Volume Exploration (DELVE) survey sky footprint.
   Subsequently, we rigorously evaluate the reliability and robustness of our estimation methodology, gauging its performance against other well-established machine learning methods to ensure the quality of our redshift estimations. Our best results constrain photometric redshifts with the bias of $-0.0013$, a scatter of $0.0293$, and an outlier fraction of $5.1\%$. These point estimates are accompanied by well-calibrated PDFs evaluated using diagnostic tools such as Probability Integral Transform and Odds distribution.   
   We also address the problem of the accessibility of PDFs in terms of disk space storage and the time demand required to generate their corresponding parameters. We present a novel Autoencoder model that reduces the size of PDF parameter arrays to one-sixth of their original length, significantly decreasing the time required for PDF generation to one-eighth of the time needed when generating PDFs directly from the magnitudes.
\end{abstract}




\end{frontmatter}




\section{Introduction}
\label{sec:introduction}



The advent of data-intensive wide-field imaging optical surveys in \colreview{astronomy}, such as the Sloan Digital Sky Survey  \citep[SDSS;][]{York2000}, the Dark Energy Survey \citep[DES;][]{Abbott2018}, the Kilo-Degree Survey \citep[KIDS;][]{deJong2013}, has been transformative in our comprehension of the Universe. These surveys facilitate the observation of large-scale structures, thereby advancing various studies in weak gravitational lensing \citep[e.g.][]{2021MNRAS.504.4312G,gavazzi2007}; galaxy evolution and morphology~\citep[e.g.][]{desi_zoo,bom2024,cid2005,fornax}; and cosmology ~\citep[e.g.][]{des_clusterdr1,cluster_mass,2024A&A...682A.148F}.

\review{ In the next years the amount of available data is expected to reach a new level with the upcoming Vera Rubin Legacy Survey of Space and Time \citep[LSST;][]{Ivezic2019}. The LSST is expected to produce $\sim 15$ PB of data and observe $\sim 20$ billion of galaxies\footnote{see \url{https://www.lsst.org/scientists/keynumbers}}.}
These extensive surveys \review{contain} a plethora of data, including the magnitudes and colors of galaxies, thereby enabling an assessment of their intrinsic properties. A \review{important} parameter obtained from these observations is the redshift of galaxies. Redshift serves as a proxy for their distances within a given cosmological model \citep{Harrison1993}. 
\review{Redshifts} are crucial for identifying objects present in wide field surveys, constraining cosmological models, and tracing the evolution of galaxy properties over cosmic time.

Traditionally, spectroscopic redshifts ($z_{\rm spec}$ ) have been regarded as the benchmark for precise redshift measurements \citep{Glazebrook1998, desidr1}. However, acquiring $z_{\rm spec}$ is both time-intensive and resource-demanding, as strong spectral features must be clearly recognized \citep{Isanto_2017}.\review{ For instance, using a single slit in a 4m-class telescope can take dozens of minutes per object at $r \gtrsim 20$}. 

Photometry-based models offer a viable alternative for extracting redshifts. Unlike spectroscopic methods, which involve resource-intensive spectroscopy \review{\citep[e.g., ][]{Bolton_2012}}, photometric redshift \review{($z_{phot}$)} estimation utilizes broadband photometry to infer galactic redshifts. \review{The latter} approach is advantageous as it can be applied to a much larger sample of galaxies, making it more scalable and cost-effective \citep{Benitez2000,eazy, sanchez2014,cipriano2023}. 

\par 
Estimating \review{$z_{phot}$} can be approached through two main methods: Template fitting and Machine Learning. Template fitting (TF) methods make use of the spectral energy distribution (SED) of an object and fit it to a template library, which can be synthetic or empirical, containing spectra of different types of galaxies at different redshifts \citep{Benitez2000,Crenshaw_2020}.  This approach necessitates highly calibrated, unbiased templates alongside explicit assumptions regarding dust extinction. Furthermore, the available template set fundamentally constrains its applicability. 

Machine Learning (ML), and particularly Deep learning (DL), has emerged as promising avenues for estimating $z_{phot}$s \citep[e.g.,][]{Newman2022, Duncan_2022}, primarily due to its capacity to learn intricate patterns within extensive datasets. 
\review{It is a viable alternative to template fitting methods once it learns a mapping from photometry directly to redshift, bypassing the need for physical assumptions and avoiding unrepresentative templates \citep{Lima_2022}.}
Additionally, utilizing massive GPU parallelization in ML offers a more computationally efficient alternative. Nonetheless, many of these ML techniques yield point estimates \citep{benchmark_2021}, which cannot thoroughly represent the additional characteristics inherent in complete Probability Density Functions (PDFs). The relevance of assessing these features has been reported for weak lensing cosmography \citep[e.g., ][]{mandelbaum2008precision, myers2009incorporating}.
Although several ML Algorithms can produce \review{PDFs \cite[][]{mucesh2021,schmidt2020evaluation, desprez2020euclid,sadeh2016annz2}}, the use of more recent probabilistic DL techniques in the context of different architectures has not yet been fully explored.
\review{In certain scenarios, PDFs from ML and DL algorithms may exhibit inconsistencies when interpreted in a probabilistic manner, necessitating careful inspections or even re-calibrations to ensure their reliability \citep{Isanto_2017, dey_2022}}. 
In this work, we explore a strategy that uses a Recurrent Neural Network (RNN), an algorithm originally designed for sequence processing
based on Legendre Memory Units \citep[LMU;][]{voelker2019lmu}, of which uses a Mixture Density Network \cite[MDN;][]{Bishop_1994} to derive PDFs. 
 We apply these methods to real data to produce a photometric redshift catalog for the DECam Local Volume Exploration Survey \review{\citep[DELVE; ][]{delvedr2}}. 
 
 \colreview{The impending start of the LSST by the Vera Rubin Observatory is anticipated to expand the paradigm of Big Data in astronomy by cataloging about 20 billion galaxies} \citep{Ivezic2019}. \review{Generating publicly available PDFs for all new galaxies discovered may result in storage management and time consumption issues. }
 \review{In this work, we also} explore the utilization of auto-encoders \review{\citep[AEs, ][]{Rumelhart_1986}} \review{to reduce the space needed to store PDF information.}
 As a demonstration of this method, we successfully compressed the entirety of \photoz's PDF information by a factor of six \review{and increased its querying speed by a factor of eight}. Embracing AEs offers a practical solution for forthcoming surveys, \review{ such as Rubin LSST, }as storing and utilizing complete redshift \review{PDFs} information for the entire observable sky would be unfeasible.

The paper is structured as follows. Section \ref{sec:data} introduces the data and outlines the various sample definitions used in the DL process. Following this, Section \ref{sec:methods} \review{explores} the DL models employed among other techniques used to estimate $z_{phot}$.  Section \ref{sec:methods} also elaborates on the models' training and validation procedures, while Section \ref{sec:results} presents the findings, including a validation of our method and comparisons \review{between it and estimators found in the literature}. Section \ref{sec:discussion}, we offer a summary and discussion of our work, along with concluding remarks. In Section \ref{sec:catalog}, we describe the \photoz catalog produced by our method \review{applied to DELVE. Finally, in Section \ref{data_acess}, we describe how to get access to the DELVE DR2 \photoz catalog.}

\section{Data}
\label{sec:data}

\subsection{DELVE Survey}
    
The DELVE survey combines exposures from the Dark Energy Camera \citep[DECam, ][]{Flaugher215} collected by devoted DELVE observing with archival data from the Dark Energy Survey (DES), the DECam Legacy Survey (DECaLS), and more than 270 community programs publicly available \citep[e.g.][]{delvedr1,delvedr2}. 
In this study, \colreview{we use DELVE data release 2 (DR2)}, encompassing an area of 21,000 deg$^{2}$\review{, of which} 17,000 deg$^{2}$ \review{has been observed} in \review{all four} broadband filters ({\it g}, \textit{r}, \textit{i}, and \textit{z}). These observations are limited \review{up to DEC $\lesssim$ 30} \textdegree, complete up to magnitudes  $24.3$, $23.9$, $23.5$, and $22.8$ for $g,r, i$, and $z$, respectively for $5\sigma$ point source detections. In this context, the DELVE-DR2 catalog comprises $2.5$ billion unique astronomical sources, of which $\sim$ 618 million \colreview{have} observations available in the four photometric bands. 

DELVE observations were processed using DESDM \cite[Dark Energy Survey Data Management,][]{morganson2018}. DESDM uses \texttt{SourceExtractor} \citep{sextractor} and \texttt{PSFEx} \citep[Point Spread Function Extractor,][]{psfex}, which provide essential parameters such as \texttt{FLAGS} and \texttt{SPREA\_DMODEL} which can be used as quality thresholds \citep{delvedr2}. In our initial steps, we use these parameters to curate a high-quality galaxy sample based on well-observed images and an extended source classification. Specifically, we exclusively considered objects with \texttt{FLAGS} $<3$, ensuring the absence of any warning signals during the detection process \citep[see Table 4 in][]{delvedr2}. In DR2, the magnitudes are not corrected for extinction. To address this, \review{we subtract} the values in columns \texttt{EXTINCTION\_{G,R,I,Z}} from the corresponding values in columns \texttt{MAG\_AUTO\_{G,R,I,Z}}, as indicated in \citet{delvedr2}. 
Additionally, we selected objects with magnitude $g < 23.5$ to avoid objects \review{near the detection limit and ensure depth homogeneity.} Finally, we used the \texttt{MODEST\_CLASS} criteria \cite[see ][]{Drlica_Wagner_2018} to remove contaminant stars by choosing the objects that lie in the classes $3$ (high-probably galaxy) and $2$ (ambiguous classification).
By applying these filters, we generated a Main Catalog (MCAT) containing 347 million sources. The MCAT constitutes the foundation for subsequent galaxy selection \review{and redshift estimation. It encompasses the input catalogs to be used training our models and all further photometric redshift estimations using \review{DL}.}




\subsection{Spectroscopic Sample}

        This investigation focuses on computing photometric redshifts using DL techniques. To accomplish this aim, obtaining accurate redshift measurements \review{(i.e., $z_{spec}$)} is essential in providing data to train the models. To construct the necessary \specz catalog with DELVE-DR2 photometric correspondences, we crossmatch \review{the MCAT with an updated version of \specz catalog described by \cite{julia2017}. This update includes new data from 
        VANDELS \citep{McLure_2018},
        MUSE \citep{Bacon_2010},
        CLASH-VLT \citep{Mercurio_2021} and 
        SDSS DR16 \citep{Ahumada_2020}
        surveys. Other data sources present in this catalog are displayed in Table A4 in \citet{julia2017}.
        
        
        Additionally, we included the early data release of the Dark Energy Spectroscopic Instrument \cite[DESI-EDR:][]{desiedr} and the Southern Hemisphere Spectroscopic Redshift Compilation \citep{erik_phz} in the ancillary dataset. 
        }
        
        
       We crossmatched MCAT and our spectroscopic catalog using a search radius of $1$ arc second. This crossmatch resulted in 2.9 million objects with reliable \review{\specz} measurements available. These 2.9 million objects constitute our ancillary spectroscopic catalog (SZCAT). \review{Since we use the MCAT to crossmatch with the spectroscopic data, SZCAT is also constrained by the initial cuts imposed on MCAT.} This catalog is used to build the truth table for training, validation, and test processes (the DL model is presented in Section \ref{sec:methods}). 


    \subsection{Data Selection}
        To avoid poor detections in our dataset, we applied additional constraints beyond the MCAT's global magnitude limit of $g<23.5$. We implement a magnitude cut of $g<22.5$ on the SZCAT which, combined with signal-to-noise ratio requirements, ensures a high proportion of reliable measurements and a dataset free from spurious sources, low-quality detections, and very faint galaxies. Additionally, we applied color cuts to exclude nonphysical objects \cite[see  ][]{Drlica_Wagner_2018}, \colreview{and we find that most of our objects have} $-0.5 \leqslant g-r \leqslant 1.5$ and $-0.5 \leqslant r -i \leqslant 0.8$ regions in the color-color space. We restricted our $z_{spec}$ interval to avoid spurious detections of galaxies located at high redshift and also to avoid stars and blueshifted galaxies at very low redshift. A summary of the photometric quality flags applied to SZCAT are shown below:
        
    
        
        \begin{itemize}
        \centering
            \item[] $-1<g-r<4$
            \item[] $-1<r-i<4$
            \item[] $-1<i-z<4$
            \item[] $griz<22.5$   
            \item [] $0.01 < z_{spec} < 1$
        \end{itemize}


    After applying all the data selection criteria, our initial SZCAT sample was reduced to 52$\%$ of its original size (see Table \ref{tab:trainABC}).
    All training and evaluation of our DL model will be performed using data from this refined dataset, which constitutes the DLCAT (Deep Learning Catalog).

\subsection{Deep Learning Sample Definitions}
\label{subsec:sample_def}
        For our problem, we utilize supervised learning \cite[see ][]{Goodfellow_2016}, where the model is trained by comparing its predictions with known results. We provide the model with \review{magnitudes and colors as input features}, and it attempts to predict the corresponding output, which in our case is \review{a PDF with the most probable value being} the spectroscopic redshift. This prediction is then compared to the actual redshift value, and the model learns from the difference to improve its future predictions.
        
        Three datasets are necessary to define all the training and evaluation processes for a standard supervised DL application. The training dataset is the one that the network uses to regress the best parameters. 
        The validation dataset, which can sometimes constitute a small subset of the training dataset, serves as the means to assess the model's optimization process during training epochs. Its primary purpose is to signal whether the network is overfitting \citep{Mohri_2018}.
        The test set is used to validate the network results for unseen data. The data comprising the test sample do not participate in any stage of the training process.

        
        
        
    The test data utilized in our study constitutes 20\% of the DLCAT dataset. Despite employing selection criteria in generating the DLCAT, we refrain from making presumptions regarding real-world observations. Consequently, no additional constraints are imposed on the test data. Subsequent paragraphs detail the composition of the training and validation datasets, underscoring that the test dataset's selection preceded any described procedures.
        
        

\begin{figure}
                \centering
                \includegraphics[width=.7\linewidth]{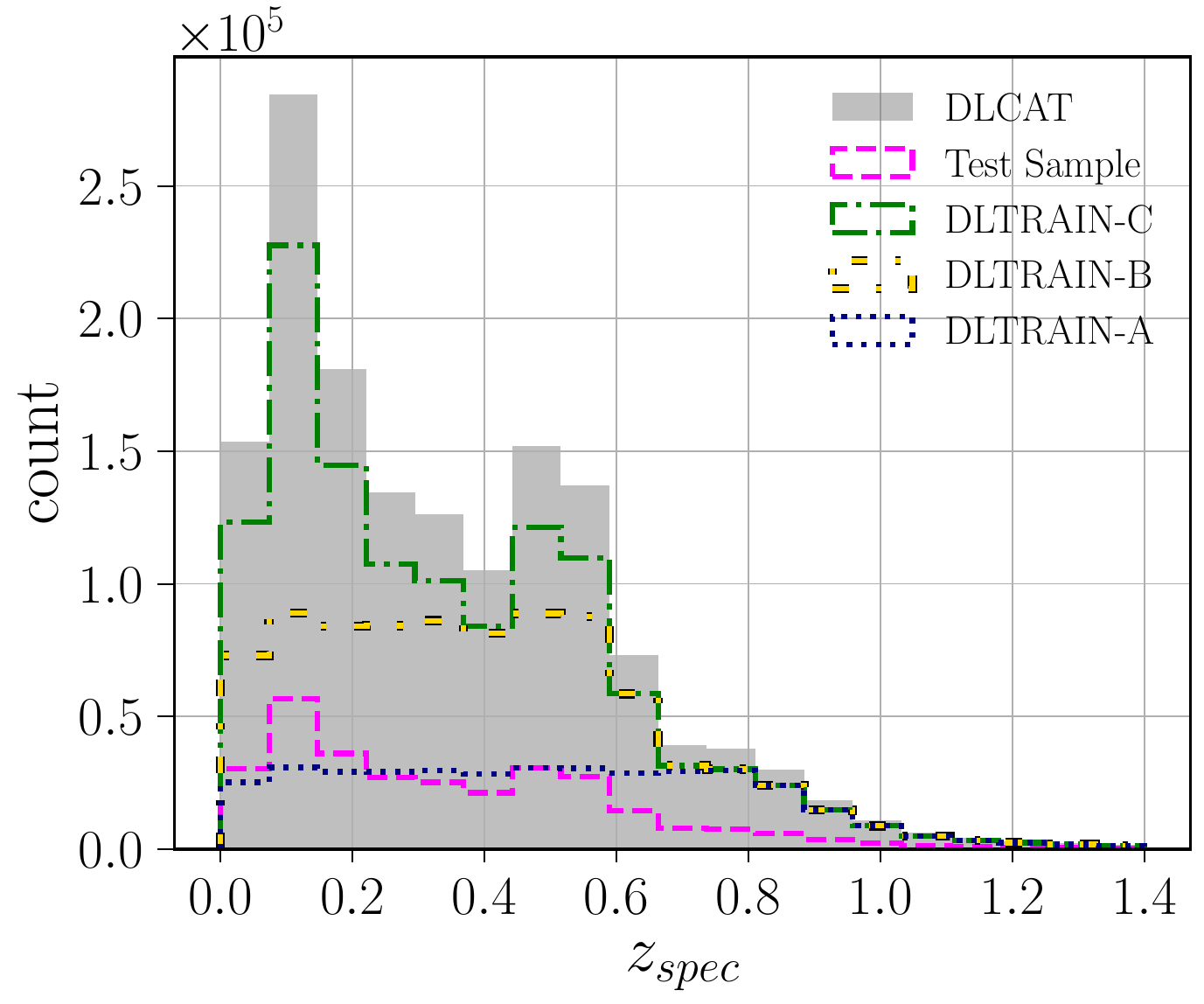}
                \caption{Spectroscopic Redshift distribution of DLCAT (grey solid shadow), DLTEST (magenta dashed line), DLTRAIN-A (navy blue dotted line), DLTRAIN-B (yellow sparse dash-dotted line), DLTRAIN-C (green dash-dotted line).}
                \label{fig:dlcat_hist}
            \end{figure}

        DL Models can be sensitive to the distribution of training data. The models usually perform better on unseen data with features similar to the ones used in the training process \citep{Jeon_2022,kim2019learning}. Therefore, the features (or target) distribution of training data plays an important role in the model's generalization ability. When dealing with supervised learning tasks, it is essential to balance the training data features as much as possible to minimize systematic biases. \review{Figure \ref{fig:dlcat_hist} displays the $z_{spec}$ distribution for the test sample, for the DLCAT, and for all Training catalogs described below.}

            \review{In the context of photometric redshift estimations, a major issue arises from the under-representation of the $z_{spec}$ distribution compared to the all-sky galaxies redshift distribution. The models may exhibit biases toward more frequent redshifts.
            Many techniques are being developed to mitigate this issue, such as Data Augmentation and the attribution of weights to loss calculations \citep{Moskowitz_2024, Schuldt_2021, limamarcos_2008}.
            
            Preliminary tests with such techniques 
            did not show significant improvement within our scope.}
            \review{Nevertheless, as an alternative approach motivated by the work of \citet{Zou_2019}, we address this issue by}
            exploring three different distributions of \specz for the training dataset. The hypothesis is that we should get less biased results with a homogeneous distribution in \specz. 
            
            We will call the training data corresponding to these distributions DLTRAIN-A, DLTRAIN-B, and DLTRAIN-C, respectively. The DLTRAIN-A consists of a homogeneous distribution from \specz $=0.01$ up to  \specz $\approx 0.8$ in bins of $0.05$ width, containing about $\approx2\times10^4$ objects per bin. 
            The DLTRAIN-B \specz distribution is similar to the \review{former} one, but homogeneous from \specz $=0.01$ up to  \specz $\approx 0.6$, keeping the same bin width and containing approximately $\approx6\times10^4$ objects per bin. 
            The main idea behind these settings for DLTRAIN-B is to quantify the impact of having more objects on the training set in contrast to the non-uniformity of the distribution towards higher values in \textit{z}. 
            

           The DLTRAIN-C comprises all the DLCAT data not present in the test sample to show via quality criteria if the distribution shape is more relevant than the amount of data to the model bias. In Table \ref{tab:trainABC}, we \review{summarize} the description of these different training distributions.

            In Section \ref{sec:results}, we show that the DLTRAIN-A is the best option for the training set to our purposes, so the reader may assume that all further results and analysis are made based on the Mixture Density Network (MDN) model trained using the DLTRAIN-A data set. A similar choice was made in \citet{Zou_2019}.
            
                
            \begin{table}
                \centering
                \begin{tabular}{|c|c|>{\centering\arraybackslash}p{0.5\linewidth}|}
                \hline Dataset & n$^o$ objects & Description
                \\\hline\hline
                MCAT & 457383847 &  All DR2 Objects with mag-\textit{g}$<23.5$, \texttt{MODEST\_CLASS}      $\in \{1,3\}$ and \texttt{FLAGS\_G}$<3$   \\\hline 

                SZCAT & 2866993 & All MCAT objects with spectroscopic correspondence through the described spectroscopic catalogs (after removing duplicates)   \\\hline

                DLCAT & 1503422 & All SZCAT objects remained from the quality cuts    \\\hline
                
                DLTRAIN-A & 384305 & Uniform distribution of  in bins of $0.05$ width from $z=0$ to $z=0.8$ containing $\approx 20000$ objects per bin; all non-Test post-selections data available for $z>0.8$ \\\hline 
                
                DLTRAIN-B & 857823 & Uniform distribution of  in bins of 0.05 width from $z=0$ to $z=0.6$ containing $\approx 60000$ objects per bin; all non-Test post-selections data available for $z>0.6$ \\\hline 

                DLTRAIN-C & 1202597 & All non-Test post-selections data available \\\hline 

                Test set & 300825 & Randomly selected 20$\%$ of DLCAT before any other selection \\\hline 
                
                \end{tabular}
                \caption{Description of the different training datasets used to quantify the generalization capacity of our model.}
            \label{tab:trainABC}
            \end{table}

            Once we have established the data sets, we use as input the magnitudes and colors (all possible colors given the available bands) of each galaxy.
            All inputs were normalized by standard deviation normalization \citep[\texttt{StandardScaler}]{2020SciPy}. 
            
            
            
            The transformation from \texttt{StandardScaler} was fitted on the training data and then applied to the test data.
            We did not apply any transformation in the \specz values.

    \subsection{Inhomogeneous Coverage}
    \label{sec:coverage_problem}
 
    The DELVE-DR2 galaxies dataset contains approximately 77$\%$  of its objects with complete photometric coverage, meaning they have observations for all four bands ({\it g}, {\it r}, {\it i}, and {\it z}). However, training a model on these four bands and their respective colors would \review{restrict us from estimating} photometric redshifts for objects missing measurements in one or more bands. Notably, only $\sim$ 26$\%$ of SZCAT objects lack full coverage, with 2$\%$ missing only the {\it z} band and 24$\%$ missing only the {\it i} band. \review{This quantity of objects is too low to provide a well-balanced training dataset for training a model specially focused on $z_{phot}$ estimation for incomplete data (e.g. with only \textit{gri} or \textit{grz} coverages).} 
    
    

    To provide redshift measurements for all objects that were observed at least in three bands, including those with more coverage ({\it g} and {\it r} bands), we trained two additional versions of the MDN model. We proceeded with the training by omitting the magnitudes {\it i} and {\it z}, respectively, from the DLTRAIN-A dataset. The performance of these models is evaluated and discussed in Section \ref{sec:results}.
    

\section{Methods}
\label{sec:methods}
    \review{
    In this work, we test several photometric redshift estimators and compare them. We propose a DL approach that converts the pre-processed photometric data (magnitudes and colors; see Section \ref{subsec:sample_def}) into redshift PDFs. Our model takes an array containing the normalized magnitudes and colors as input, processes this data, and generates the parameters that define the PDFs, modeled as linear combinations of Gaussians. The training consists of determining the output parameter set that maximizes the PDF at the spectroscopic redshifts of each galaxy.
    
    Additionally, we propose using a DL-based technique that compresses the size of the parameter space, efficiently optimizing the PDF storage. This section provides an overview of deep learning models and architectures, detailing their functionality and applications. We also describe the benchmark techniques from the literature we adopted for comparison purposes and discuss the adaptations made to these techniques for their application to DELVE data.
    
    }


    

    

    \subsection{Deep Learning }
    
        
        \subsubsection{Legendre Memory Units}
            
            Recurrent Neural Networks (RNNs) are an architecture adapted to work with ordered data by processing every input step sequentially. The backpropagation also runs through the input sequentially, which could give rise to the vanishing or exploding gradient problem. Furthermore, {vanilla RNNs} have low memory capacity; they cannot retain information from more than a few timesteps back. 
            \par \textit{Gated} RNNs, such as the Long-Short Time Memory \citep[LSTM,][]{lstm} and the Gated Recurrent Unit \citep[GRU,][]{gru} were created to solve or at least mitigate these problems. These cells use a memory state across timesteps and have gates to control the flow of information, learning what is to be kept and what can be discarded. 
            However, natural processes often have a continuous time flow, and it is unclear how LSTMs can perform in these situations, as the memory capacity required is very large.
            \review{\par Legendre Memory Units \citep[][LMU]{voelker2019lmu} are a type of recurrent cell able to learn long-range dependencies even as the input is nearly continuous, by orthogonalizing the time dimension of the input across a sliding window of a given length through the use of Legendre polynomials.

            \begin{figure}
                \centering
                \includegraphics[width=\linewidth]{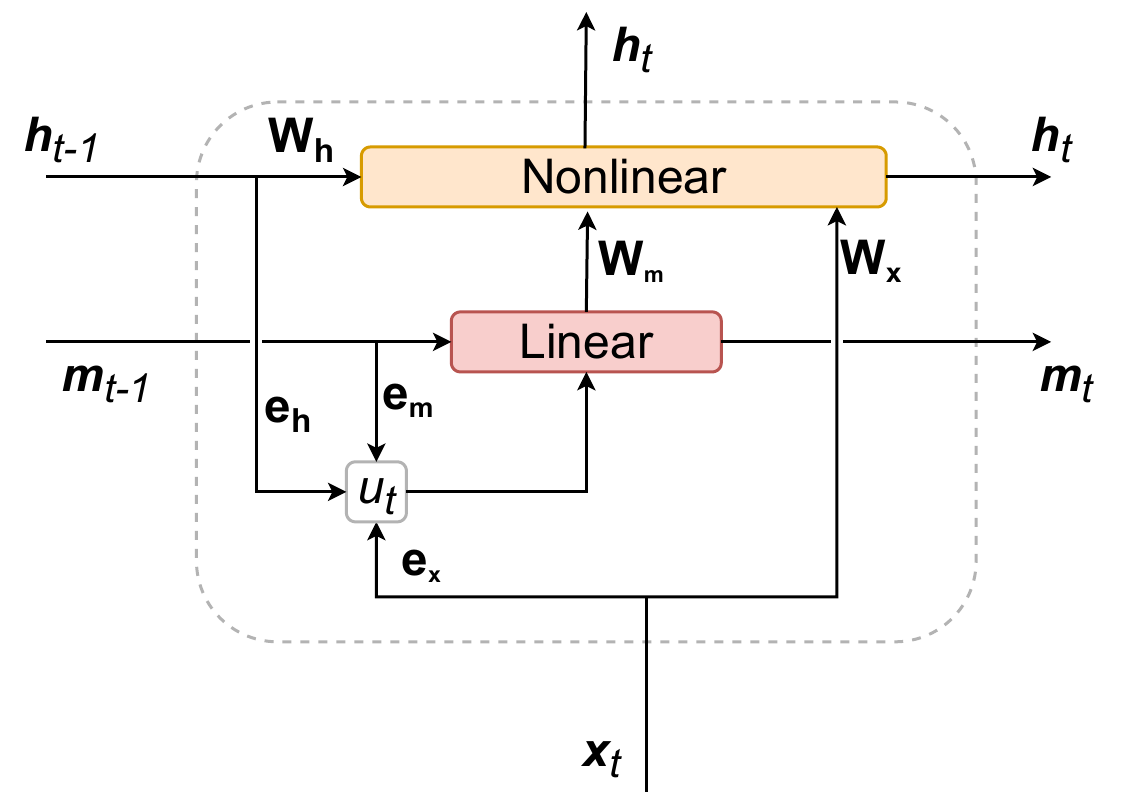}
                \caption{Structure of the LMU cell. Adapted from \citet{voelker2019lmu}.}
                \label{fig:lmu_cell}
            \end{figure}

            \par The computational graph of the LMU cell is presented in Figure \ref{fig:lmu_cell}. It couples a Linear Time-Invariant system (LTI) and a nonlinear one. The linear system updates the memory vector (\textbf{m$_t$}), orthogonalizing the input into sliding windows. In contrast, the nonlinear system uses this memory to learn arbitrary functions across time (the so-called hidden state \textbf{h$_t$}). The encoding vectors used to build the input to the linear system (\textbf{e$_x$}, \textbf{e$_m$} and \textbf{e$_h$}) and the weight matrices used for the nonlinear system (\textbf{W$_x$}, \textbf{W$_m$} and \textbf{W$_h$}) can be optimized via backpropagation.
            }
            \par For some tasks, the LMU has been shown to outperform similarly-sized LSTMs while significantly reducing training and inference times and having higher memory capacity \citep{voelker2019lmu}.
            
        
        \subsubsection{Mixture Density Networks}
        \label{sec:mdn}

        Mixture Density Networks \citep[MDN; ][]{Bishop_1994}, are a type of model combining a neural network and a (parametric) mixture model \citep{McLachlan_2019, goushen_2010}. Unlike traditional neural networks, which are trained to predict the desired parameter's single value, an MDN's output is a conditional probability density built by a linear combination of several individual probability distributions (kernels). This way, a more complete characterization of the predictions, including assessing their errors, is possible. The neural network outputs the parameters and the weights (called \textit{mixing coefficients}) of each distribution (conditioned on the inputs), which are then combined to build the mixture model. In principle, a mixture model is flexible enough to model any arbitrary distribution, making MDNs a powerful technique for regression problems. 
        
        \par Different models have been used to obtain \photoz distributions, including template-fitting \review{\citep[BPZ, ][]{Benitez_2000}} and machine learning methods \citep[ANNz2][]{cavuoti_2015}. \citet{Lima_2022} has shown that an MDN with a Gaussian Mixture Model \citep{reynolds2009gaussian} outperformed both BPZ and ANNz when inferring \photoz PDFs with SPLUS data. We then employ a similar model in this work such that the resulting PDF is given by a linear combination of $C$ Gaussian kernels, with \textit{mixting coeficients} $\{\alpha_k\}$, means $\{\mu_k\}$ and standard deviations $\{\sigma_k\}$:
        \begin{equation}
            \label{eq:mdn_final_pdf}
                PDF(z) = \sum^C_{i=1} \alpha_i \mathcal{N}(\mu_i, \sigma_i),
            \end{equation}
            with $\sum^C_{i=1}\alpha_i=1$ and $0<\alpha_i<1 $.
        \par
\begin{figure*}
            \centering
            \includegraphics[width=\linewidth]{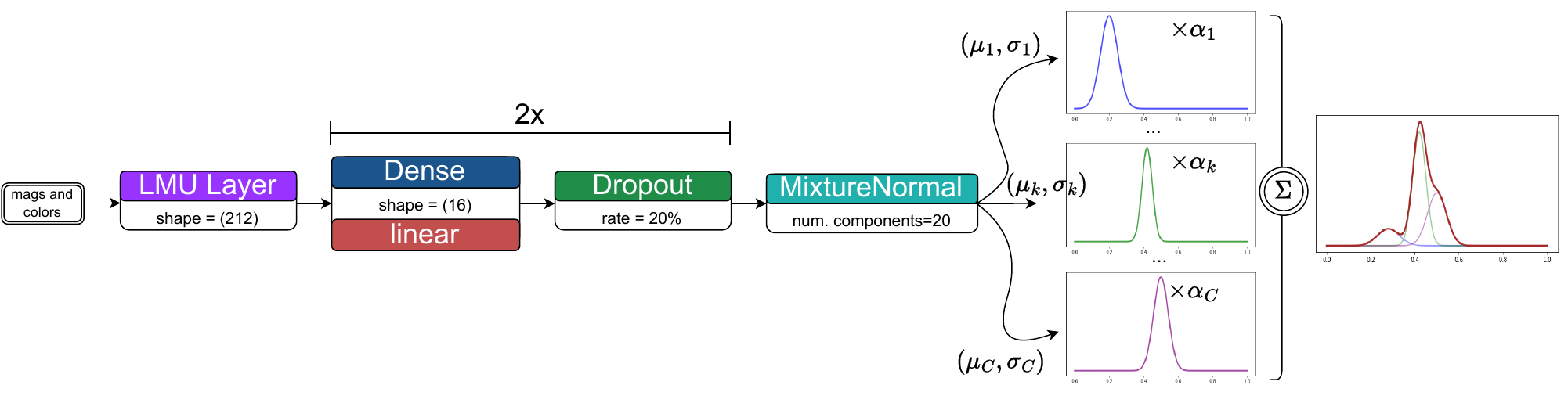}
            \caption{MDN architecture diagram.}
            \label{fig:mdn_diagram}
        \end{figure*}

 We tested all architectures developed by our team for The LSST-DESC 3x2pt Tomography Optimization Challenge \citep{Zuntz_2021}, and the one based on LMU layers had the best performance on DELVE-DR2 for \photoz regression. The architecture is shown diagrammatically in Figure \ref{fig:mdn_diagram}. It consists of an LMU layer with 212 units, followed by two fully connected layers with $20\%$ dropout each, and a mixture model of 20 Gaussians (implemented as a \texttt{MixtureNormal} layer from \texttt{tensorflow}). The network was implemented using Tensorflow 2 and Tensorflow Probability \footnote{Tensorflow v2.9.1; Tensorflow Probability v0.17.0; keras-lmu v0.5.0} \citep{tensorflow2015-whitepaper}.

        \par
\begin{figure}
            \centering
            \includegraphics[width=.85\linewidth]{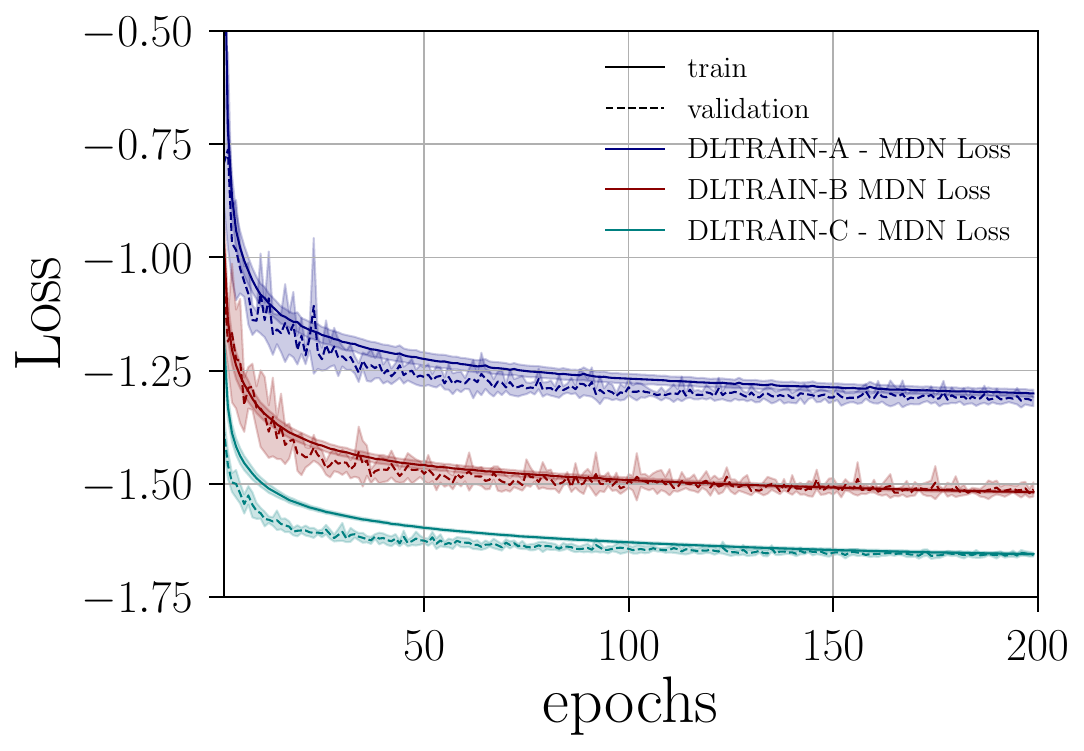}
            \caption{Training curves for the MDN models trained on the DLTRAIN -A, -B and -C datasets using the GRIZ coverage. }
            \label{fig:abc_losses}
        \end{figure}

 We trained our model with all training sets mentioned in Table\ref{tab:trainABC} and Section \ref{subsec:sample_def}. To assess the robustness of the model, we employed a k-fold cross-validation: the training set is divided into $k$ groups, and we train $k$ different models from scratch, selecting one of the groups for validation and $k-1$ for training. \review{Since our output is a PDF computed directly by the \texttt{MixtureNormal}, we choose our loss to be the negative of the PDF value at the spectroscopic redshift. In that sense, we train our model to give PDFs with the $z_{spec}$ being the most probable value.}
        We train the models for 200 epochs for each fold and select the model with the lowest validation loss at each fold to perform the predictions. \review{Figure \ref{fig:abc_losses} shows the convergence of the model training for DLTRAIN -A, -B, and -C  training sets using the full coverage input.}

        \par
\begin{figure}
            \centering
            \includegraphics[width=.85\linewidth]{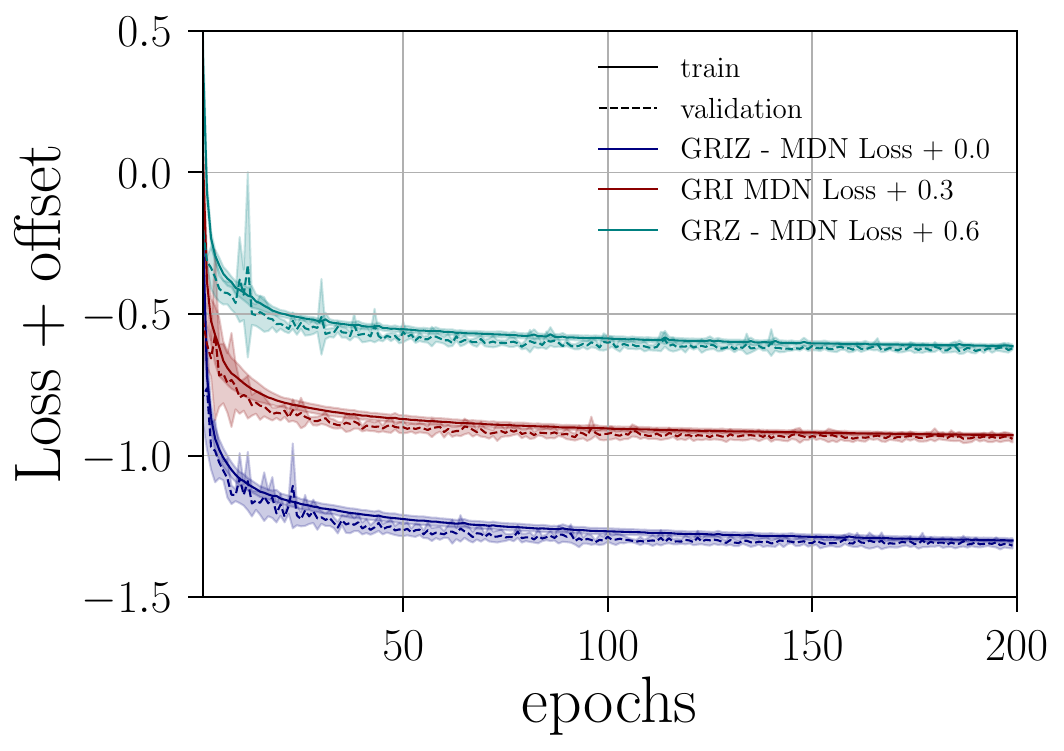}
            \caption{Training curves for the MDN models trained on DLTRAIN-A with differents coverage combinations. }
            \label{fig:coverage_losses}
        \end{figure}

 The full coverage datasets were trained on two NVIDIA RTX 3090; the training took 13s/epoch, 26s/epoch, and 35s/epoch for DLTRAIN-A, -B, and -C, respectively. The $gri$ and $grz$ versions were trained in one GPU, taking approximately 10s/epoch. Figure \ref{fig:coverage_losses} shows the convergence of the model for the DLTRAIN-A set for each coverage configuration. 

       \review{For both Figures \ref{fig:abc_losses} and \ref{fig:coverage_losses}, the shaded areas correspond to the standard deviation across the multiple folds resulting from the cross-validation process. Figures presented in Section \ref{sec:results} follow the same logic.}

        \subsubsection{Auto-Encoder}
            Auto-encoders  \cite[AEs,][]{Rumelhart_1986} are neural networks architectures first designed to dimensionality reduction tasks. 
            \review{Applications of AE include} image compressing \citep{Feng2024}, signal denoising \citep{Lin2023}, natural language processing \review{\citep[NLP, ][]{Shankar2022}} among others. \review{NLP} also provide useful tools for reducing data storage requirements \citep{Berahmand2024, liu2020multiresolution, Feng2024}. 
            The architecture of an AE comprises two parts: an encoder and a decoder.
            The encoder transforms the data to a latent space of lower dimensionality, while the decoder converts back the latent space vectors to the original size\citep{Goodfellow_2016,Wang_2014}.

            

            The AE applications are crucial because the best-case scenario is to store and use the complete redshift PDF information. But it would be unfeasible to store all this information for the entire observable sky. As described in Section \ref{sec:mdn}, each PDF is described as a set of 60 parameters, which means that storing the PDFs onsists of storing this parameter. 
            
\begin{figure}
                \centering
                \includegraphics[width=.6\linewidth]{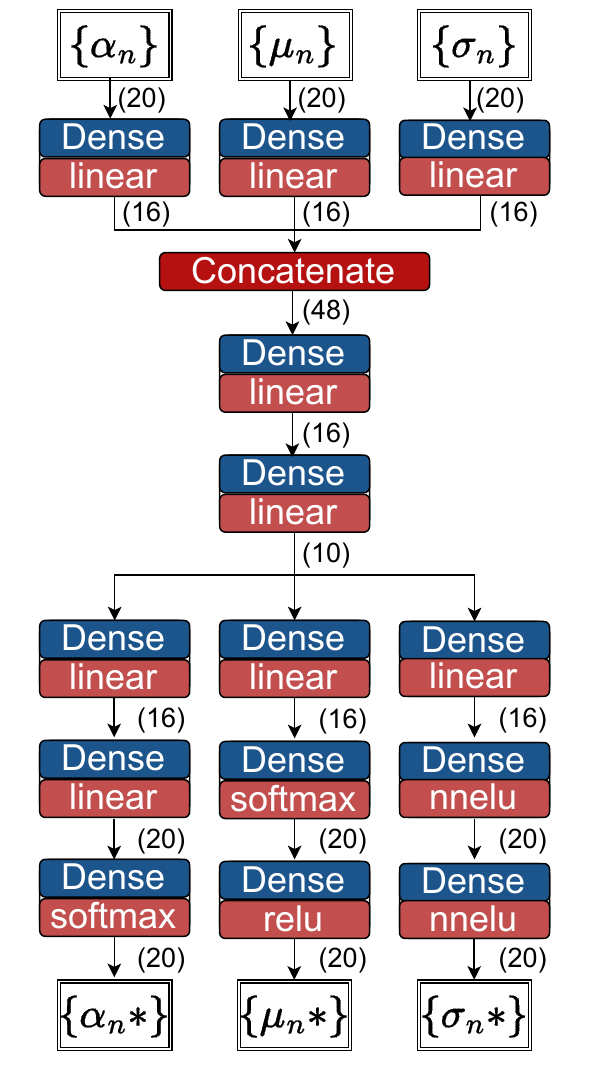}
                \caption{Autoencoder architecture diagram.}
                \label{fig:autoencoder_diagram}
            \end{figure}

            In this work, we explore the capability of an AE to compress the entire \photoz's PDF information.  Our network is trained to convert the 60 parameters predicted by the MDN model to a latent space of size 10 and then transform this 10-dimension vector again to the PDF parameters. The AE architecture used in this work is shown in Figure \ref{fig:autoencoder_diagram}. All the activation functions and hidden layers are the conventional ones defined in the Tensorflow API, except for the Non-Negative Exponential Linear Unit (\texttt{nnelu}), which modifies the \texttt{elu} \citep{elu} by adding $1$ to avoid negative values for $\sigma$. Several combinations of activation functions, number of layers, and number of neurons in each layer were tested, but this one showed the most satisfying results. We aim to apply hyperparameter tuning \review{\citep[e.g. ][]{JMLR:v24:20-1355}} in this model in the future.       

            This architecture was trained on the PDFs generated by the MDN trained on DLTRAIN-A using the total coverage information. The model was trained for 200 epochs on one NVIDIA RTX 3090, taking 4s/epoch.

        \subsection{Benchmarks}
        \label{sec:benchmark}
        \subsubsection{Machine Learning Benchmarks}
            The MDN model described \review{in Section \ref{sec:mdn}} is a non-standard DL technique that combines a recurrent neural network (LMU layer), traditional Multi-Layer Perceptrons \cite[or Dense Layers,][]{Goodfellow_2016} and a probability density output. In order to compare its performance with standard machine learning methods for photometric-redshifts, we trained a \textit{Random Forest} \cite[RF,][]{Breiman2001} model and a \textit{Fully Connected MLP} model using the DLTRAIN-A. \review{\citet{benchmark_2021}} presents a systematic analysis of several machine learning methods for estimating photometric redshifts, including the RF and the MLP. They conclude that the RF had the best performance in terms of quality metrics, \review{even though it had} the slowest training process.
            We adapted the parameterization of the RF and MLP models from the benchmarking ones derived from their work.
            The RF model was built in the Scikit-learn API\footnote{sklearn v1.1.2} \citep{pedregosa2011scikit} and the MLP in Tensorflow (same version used to the MDN).
        
        \subsubsection{Bayesian Photometric Redshifts}
            Bayesian Photometric Redshifts (BPZ) is a Bayesian template-fitting code described in \citep{Benitez_2000}. BPZ works by combining a maximum likelihood method with priors based on a set of spectra. In short, it generates Gaussian likelihoods for all spectra in the template set, then convolves each likelihood with each spectrum, and picks the maximum a posteriori redshift.

            BPZ is a vital benchmark for comparing the results of this paper, as it does not rely on the training sample used by the machine learning methods described earlier. Furthermore, it has been used in surveys such as S-PLUS \citep{molino_2020} and DES Y1 \citep{2018MNRAS.478..592H} for studies of large-scale structure.
            
            The configuration we use for BPZ is similar to that of DES Y1 \citep{2018MNRAS.478..592H}. Specifically, we use the same set of 6 templates and DES Science Verification 1 (SVA1) priors, described in \review{Section} 4.1.9 of \citep{10.1093/mnras/stu1836}. Further, we use Source Extractor automatic $griz$ photometry and assume a photometric zero point error in BPZ of 0.02.

\section{Results}
\label{sec:results}
    The Results section begins with an analysis of the accuracy of our redshift estimation model in retrieving spectroscopic redshift values. This evaluation is conducted by assessing point statistic metrics, as detailed in Section \ref{subsec:point_results} . Following this, in Section  \ref{subsec:pdf_anal} we employ various diagnostic tools to assess the quality and reliability of the \review{PDFs} generated by our model.

    \subsection{Point Statistics analysis}
    \label{subsec:point_results}
        
        We call point statistics the computation of the metrics that compare the \photoz values predicted by the network with the spec-z values. The $z_{phot}$ is the redshift value corresponding to the PDF's most probable value.

        To assess the quality of the point estimations, we compute the \textit{Normalized Median Absolute Deviation}, $\sigma_{NMAD}$, of the bias $\Delta z = z_{phot}-z_{spec}$. Following the definition of \citep{Brammer_2008} and \citep{Lima_2022}.   $\sigma_{NMAD}$ is defined as :

        \begin{equation}
            \sigma_{NMAD} = 1.48\times median\left(\left|\frac{\Delta z - median(\Delta z)}{1+z_{spec}}\right|\right)   
        \end{equation}
        
        Different from the standard definition of $\sigma_{NMAD}$ \citep{Ilbert_2006} or \citep{Li_2022}, the definition used here is less sensitive to outliers. In the present work, we call outliers (also known as catastrophic errors \citep{Ilbert_2006}) \review{the galaxies that satisfy} $\frac{\Delta z }{1+z_{spec}} > 0.15$. For any subsample of our datasets, we define $\eta$ as being the fraction of outliers on this subsample. The choice of the threshold 0.15 for outliers used here follows the one adopted in \citep{Ilbert_2006}. 
        An alternative way to define outliers is  in terms of $\sigma_{NMAD}$. \citet{Brammer_2008} uses  $5 \times \sigma_{NMAD}$ as the outlier threshold, which would correspond to $\approx 0.15$ for our entire test sample in terms of the results given by the MDN method.  

\begin{figure*}
            \centering
            \includegraphics[width=.85\linewidth]{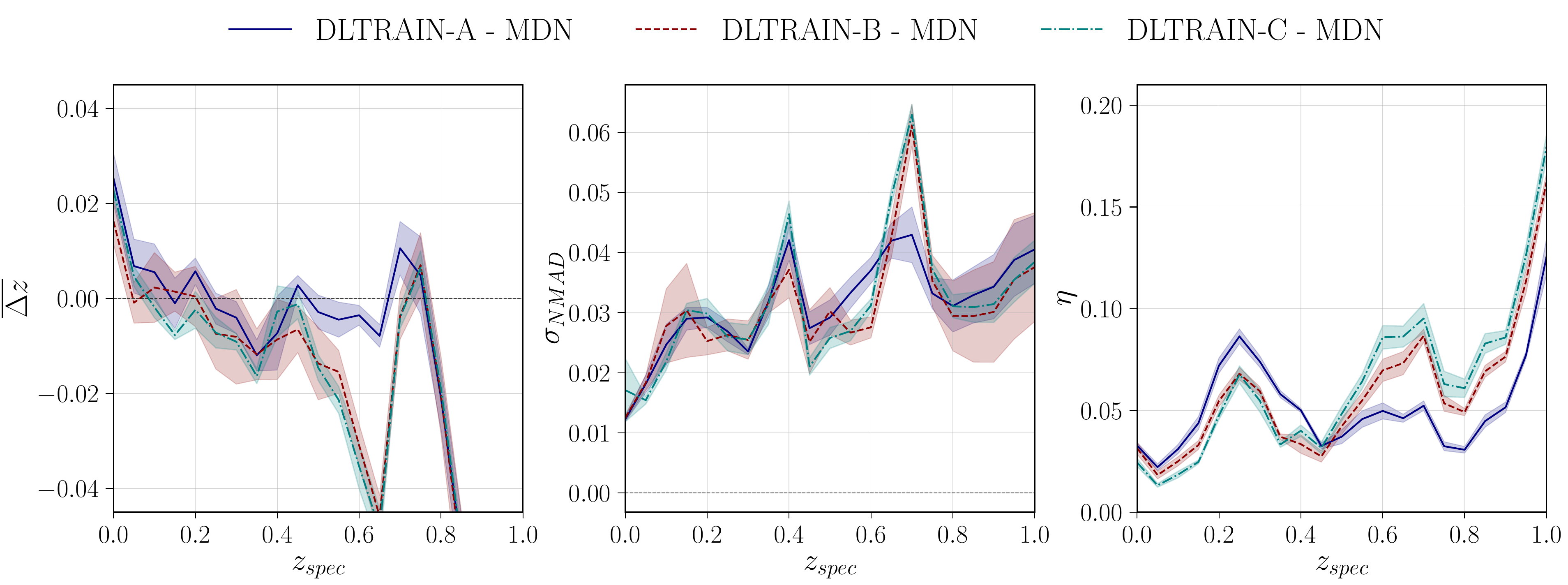}
            \caption{Point-Metrics results as function of $z_{spec}$ for the three different training sets using our MDN model. Solid lines represents the mean values of the metrics in each bin between all folds. The shaded regions represents the respective standard deviations.}
            \label{fig:abc_metrics}
        \end{figure*}

\begin{figure*}
            \centering
            \includegraphics[width=.85\linewidth]{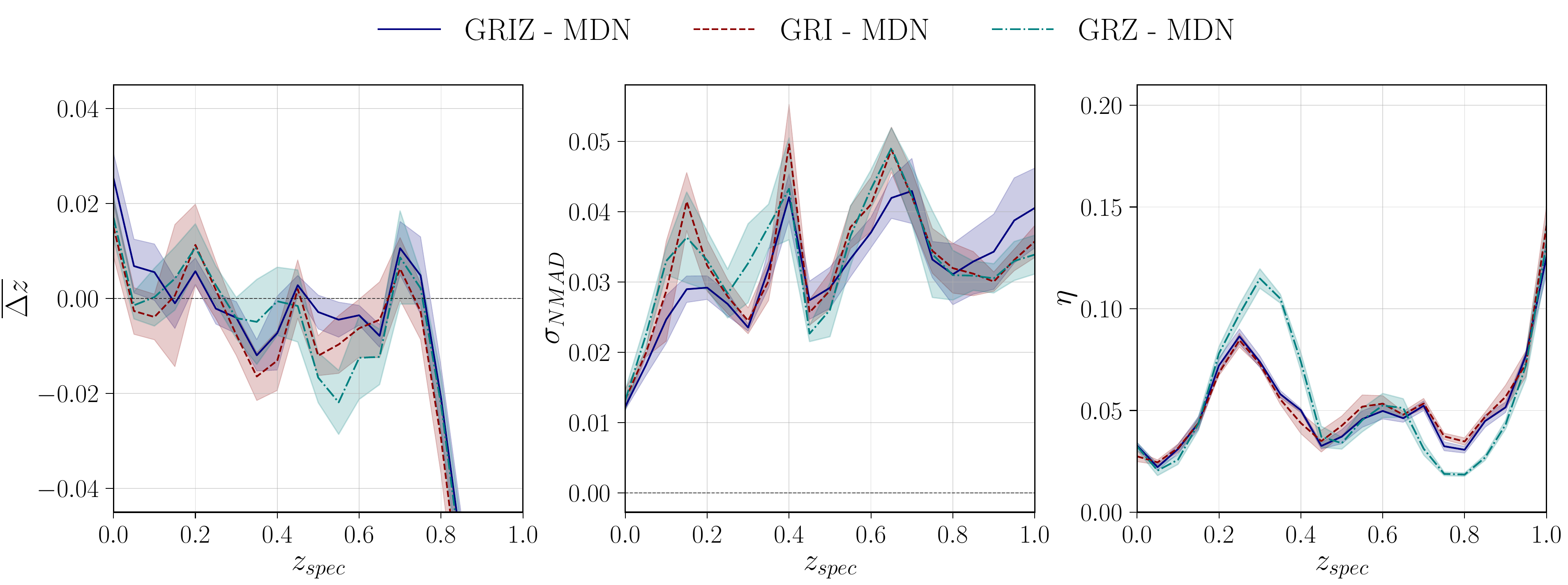}
            \caption{Point-Metrics results as function of $z_{spec}$ for the three different coverage configuration using our MDN model on DLTRAIN-A training set. Solid lines represents the mean values of the metrics in each bin between all folds. The shaded regions represents the respective standard deviations.} 
            \label{fig:coverage_metrics}
        \end{figure*}

\begin{figure*}
            \centering
            \includegraphics[width=.85\linewidth]{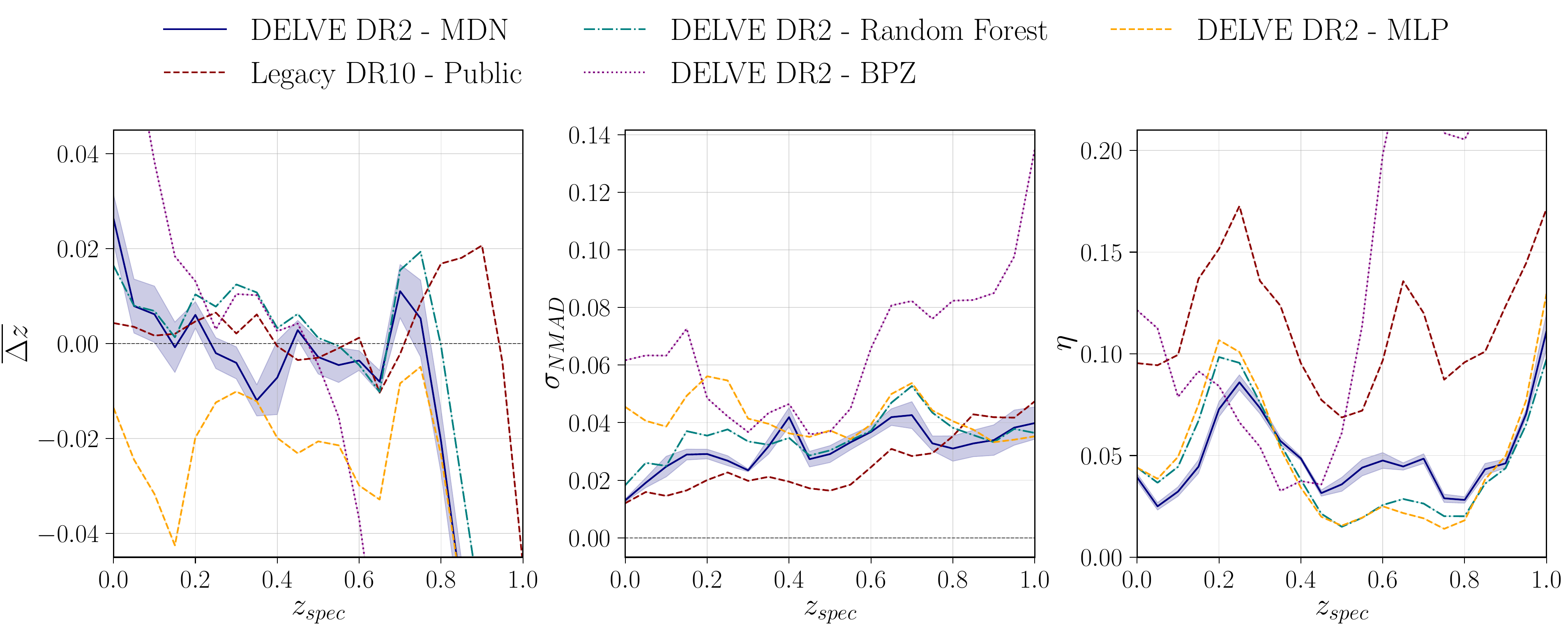}
            \caption{Point-Metrics results as function of $z_{spec}$ for the three different methods used on the test set. On the blue solid lines are the results for the DELVE DR2 data using our MDN model; \review{the teal dot dashed lines, yellow dashed lines and magenta dotted lines shows the results for the same data, although using the random forest, MLP and BPZ methods described described in Section \ref{sec:benchmark}, respectively; the darkred dashed lines correspond to the publicly available $z_{phot}$ on the LS-DR10 catalog. Our method has a lower scatter than all methods besides LS-DR10, but our catastrophic outlier fraction is much lower than theirs.}}
            \label{fig:metricsXzspec}
        \end{figure*}

\begin{figure*}
            \centering
            \includegraphics[width=.85\linewidth]{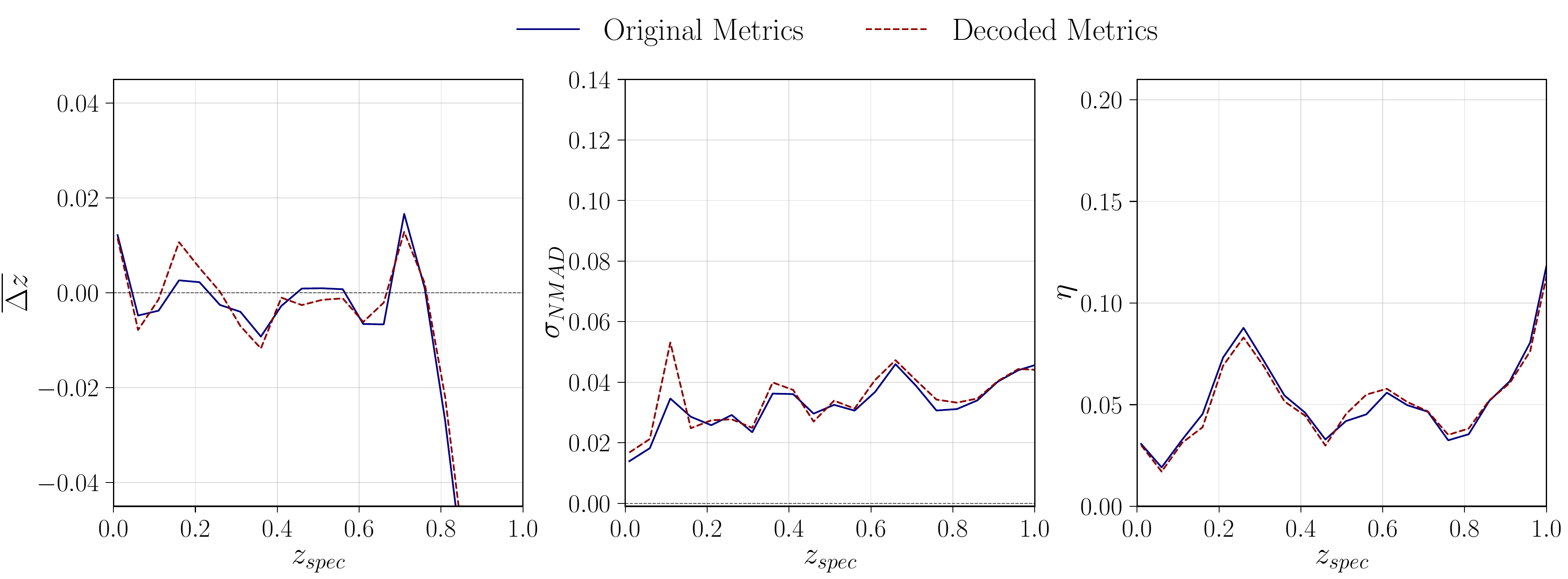}
            \caption{Point-Metrics results as function of $z_{spec}$. On the blue solid lines are the results for the DELVE DR2 directly from our MDN model. The darkred dashed lines correspond to the results that came from the reconstructed PDFs given by the Decoder part of our Autoencoder Model.}
            \label{fig:autoencoder_metrics}
        \end{figure*}

        Figures \ref{fig:abc_metrics}, \ref{fig:coverage_metrics}, \ref{fig:metricsXzspec} and \ref{fig:autoencoder_metrics} present the point-estimate results for all discussed methods applied on the test dataset. We considered all the 5 folds computing the metrics from the MDN's \photoz. The solid lines and the shaded areas represent the mean and standard deviation of the metrics through all folds, respectively. 
        
        In Figure \ref{fig:abc_metrics}, we compare the performance of the MDN model trained on the datasets DLTRAIN -A, -B, and -C using all the magnitudes.
        The graphs show that the DLTRAIN -B and -C training causes a pronunciation of bias around $z_{spec}=0.65$. The performance of the A-trained MDN is better than the other two in terms of bias, \review{except in the region \specz$<.15$, where the B- and C- trained approximate the $\overline{\Delta z}=0$ line better than MDN.} Considering the error bars, there is no relevant influence of the different training set choices in terms of scatter apart from the region around $0.6$. We can also see a \review{slightly higher} presence of outliers for the A-trained MDN relative to the other training choices up to \photoz$\approx0.46$. However, for higher $z$ the A-trained MDN shows the best performance. 

        We present the effect of missing bands in Figure \ref{fig:coverage_metrics}. We trained the MDN model using the DLTRAIN-A data in the $griz$, $gri$, and $grz$ configurations discussed in Section \ref{sec:coverage_problem}.
        
\begin{figure*}
            \centering
            \includegraphics[width=\linewidth]{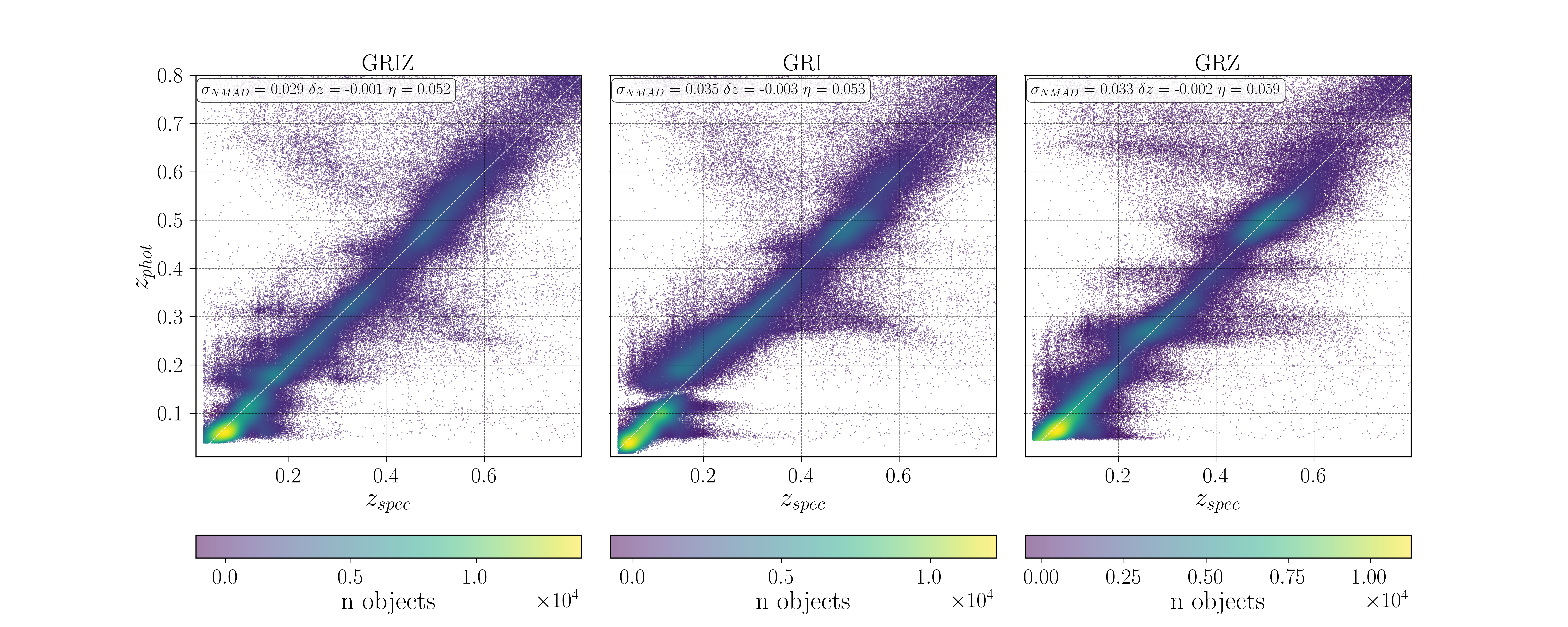}
            \caption{Distribution of $z_{phot}$ predictions versus $z_{spec}$ for the models trained in different coverages using DLTRAIN-A. The color map indicates the density of objects and we display the metrics computed for the entire test dataset on the top of each panel.} 
            \label{fig:density_cov}
        \end{figure*}

        All the cases exhibit similar bias, except in the region \specz$<0.15$, where the models trained with missing bands display a higher bias. This behavior is intriguing as it suggests improved performance in a region where less informative data contributes. Further discussion on this aspect is provided in Section \ref{sec:discussion}.
        The model trained in $griz$ magnitudes generally shows a lower scatter. It is worth noting that we are considering the mean and standard deviation between all folds. This means that \review{when} focusing on the best fold alone, the model could yield an even better curve for the scatter with full coverage.
        The $grz$ trained model exhibits an anomalous peak in the outlier fraction near \review{\specz$=0.3$}, but it shows a lower presence of outliers for $z_{spec}>0.65$. There is no remarkable distinction between the $griz$ and $gri$ trained models regarding outlier fraction. All these features are also shown in Figure \ref{fig:density_cov}, which displays the distribution of our predictions versus $z_{spec}$. \review{From this figure, we can see the $grz$ and $gri$ trained models outperforming in $z_{spec}<0.15$, however, the $gri$ trained model clearly presents a bias in the region $0.15<z_{spec}<0.3$ and the $gri$ one shows a higher bias and dispersion at the top region of the panel.} 
    
        Figure \ref{fig:metricsXzspec} presents our point estimate results for all methods described in Section \ref{sec:methods}. \review{We also computed the point-like metrics using the photometric redshift values available for the same galaxies in the public LS-DR10 catalog \footnote{https://www.legacysurvey.org/dr10/}. The resulting curves related to the LS-DR10 data are shown in Figure \ref{fig:metricsXzspec}. It is worth mentioning that when crossmatching the test sample with the LS-DR10 catalogs, a few objects were missed. We ensured these objects were removed from the test sample to compute the curves for the metrics in this figure. However, this did not affect the results significantly. }
        \review{To ensure the consistency of our proposed approach, we employed cross-validation for all the MDN-based methods. During our analysis, the other models did not exhibit any signs of overfitting or training-related issues. Therefore, we retained a single model for each comparison analysis.}
        The other methods outperform the BPZ and MLP in terms of bias and scatter. The BPZ model exhibits a noticeable bias divergence beyond $z=0.6$, recalling the behavior observed in the B- and C- trained models, as depicted in Figure \ref{fig:abc_metrics}. However, unlike the latter models, the BPZ model does not present improvements in bias after the observed divergence. 
        \review{Our method shows a valley in bias around \specz$=0.3$, different from the RF, LS-DR10 and BPZ results. However, in this region,  RF and BPZ are distinguishable only in terms of their signal, as they present similar values in the modules.}
        The LS-DR10 photometric redshifts in the general picture show a lower scatter than the other models. The MDN model outperforms the RF in terms of scatter. The LS-DR10 \photoz has a higher outlier fraction than the other models. Our model also outperforms the RF and MLP methods up to \specz$\approx0.35$, but it shows a raised presence of outliers at higher \specz.

        In order to validate the PDF's codification by the AE, we compare the results given by the Original PDFs (those that came directly from the MDN model) with the Decoded PDFs (those that came from the AE decodification). Since the autoencoder was trained without cross-validation, no error bars are associated with this analysis. It is important to highlight that this approach only ensures comparability among models trained with identical datasets. Figure \ref{fig:autoencoder_metrics} shows the point metrics for the MDN model and the AE-decoded PDFs. \review{Although the curves present small deviations from each other, considering the order of magnitude of the metric errors presented in Figures \ref{fig:abc_metrics}, \ref{fig:coverage_metrics}, and\ref{fig:metricsXzspec}, the three metrics are visually compatible.} 
    
        \review{All the point-estimate metrics are displaced in Table \ref{tab:metric_results}. The * symbol refers to computations where the test sample was slightly modified by removing objects missed in the LS-DR10 crossmatch.} 

    \subsection{Probability Density Function analysis}
    \label{subsec:pdf_anal}
        To evaluate the quality of the \photoz's estimations, we compared them to their respective target values (\specz). When working with PDF's estimation, defining the target for the PDF is not trivial. During training, the model aims to maximize each object's probability of \specz. 
        The quality of the PDFs has to be defined in terms of their features.  We first evaluate the PDFs using two different metrics: the Probability Integral Transform \review{\citep[PIT; proposed by]{dawid_1984}} and the Odds distributions.
        
        For each galaxy, the PIT can be defined as the value of the Cumulative Density Function (CDF) at its respective \specz \citep{Lima_2022, polsterer_2016, Mucesh_2021}

        \begin{equation}
            PIT = \int^{z_{spec}}_{-\infty}PDF(z)dz
        \end{equation}

        The meaning of the PIT comes from its distribution for the whole or any reasonably large sample of galaxies. As discussed by \citet{Mucesh_2021}, well-calibrated PDFs should produce a uniform distribution of the PIT values.\review{ This feautre} arises because, in principle, the PDFs are not correlated. Consequently, the CDF's values for the \specz must follow a uniform distribution $U([0,1])$. 
        
        A PIT distribution's shape deviating from a uniform distribution informs about possible systematic behaviors of the inferred distributions. When concave or convex, the PDFs tend to be underdispersed or overdispersed, respectively. Moreover, the presence of slopes on the distribution indicates biased estimations \citep[see][]{polsterer_2016}.

       The Odds, originally introduced as Bayesian Redshifts Odds  \citep{benitez_2014jpas}, measure the degree of reliability for a given PDF. It is defined as the probability of redshift in an interval around the \photoz. For a given PDF, we can compute
        
        \begin{equation}
        \label{eq:odds}
            Odds = \int^{z_{peak}+\xi_z}_{{z_{peak}-\xi_z}}PDF(z)dz
        \end{equation}

        We wrote the integration (Eq. \ref{eq:odds}) in terms of the PDF most probable value $z_{peak}$ \review{--- which is also $z_{phot}$ in our definition} \citep{Lima_2022, Benitez_2000}. 
        If the $Odds$ has a high value, the PDF is precise around $z_{peak}$. Therefore, a low value of $Odds$ can be interpreted as a broad PDF, thus making it inaccurate.
        \citet{Lima_2022} uses the value of $\xi_z=0.02$ to compare the deviation in photometric redshifts PDFs encountered in \citet{molino_2020} using the BPZ. \citet{Coe_2006} uses the value of $0.06$ to analyze BPZ results for galaxies in the Hubble Ultra Deep Field \review{\citep{beckwith2006hubble}}. We adopt the value of $\xi_z=0.06$ for this work. 
        Confident and reliable PDFs produce a $Odds$'s distribution with a peak close to one.    

\begin{figure}
            \centering
            \includegraphics[width=\linewidth]{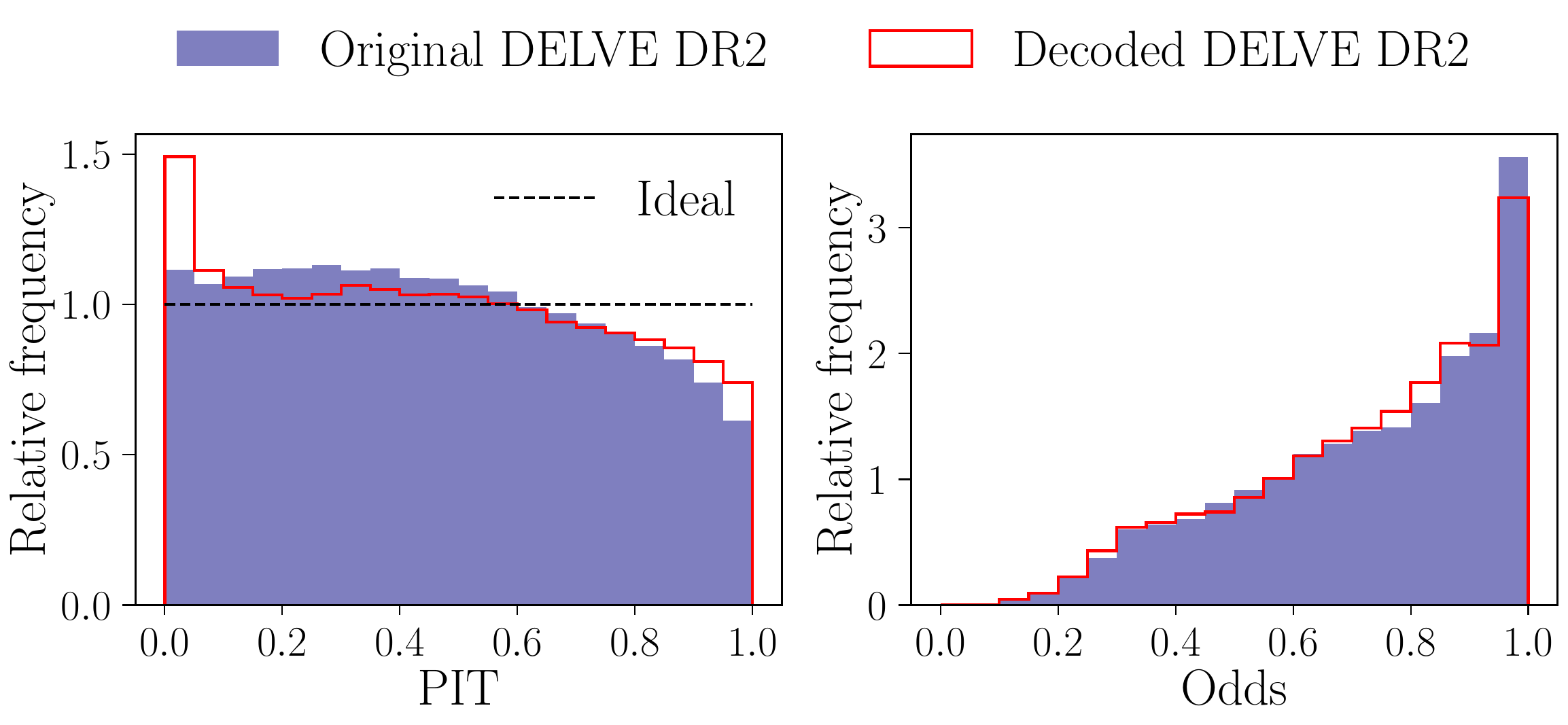}
            \caption{PIT and ODDS distributions. The blue solid bars represent the relative frequency of the metrics calculated from the original MDN PDFs prediction for the test dataset. The red lines in steps show the respective distributions for the reconstructed PDFs from the Decoder part of our Autoencoder Model.}
            \label{fig:PIT_ODDS}
        \end{figure}

        Figure \ref{fig:PIT_ODDS} shows the PIT and Odds distribution for the PDFs from the MDN model and its decoded version from the AE. The PIT distribution exhibits a \review{shallow negative} slope for the MDN model, which indicates that the model \review{may be} underestimating the PDF peaks values \review{of the PDF} relative to the spec-z values. For the Odds distribution, the peak is close to one, as expected. Therefore, our model tends to produce precise PDFs. 
        The decoded PDFs present an anomalous peak in $PIT\approx 0$ \review{and higher frequencies relative to the original PIT values at $PIT\approx 1$. We further discuss this }in \ref{sec:discussion}. Apart from this anomaly, the decoded PDFs produce PIT and Odds distributions similar to the original ones.

        \subsubsection{Coverage Test}

         The coverage test (CV-Test or Coverage diagnostic) is a well-known metric usually used to discuss the quality of the credible regions produced by simulation-based inference algorithms \citep{hermans2020likelihoodfree, hermans2022trust}, which can also offer a way to evaluate the quality of the estimated redshift PDFs as described in \citet{Dalmasso_2020}. Given the observation and an estimated distribution, one can assess the probability that a certain credible region of the inferred distribution contains the true value. This gives an idea of whether the estimated distribution is overconfident, calibrated, or underconfident.
        
       The CV-Test was made by taking the pair (\specz, PDF) and sampling 1000 times from the \photoz's PDF, producing a frequency distribution with 2000 bins ranging from 0 to 2. 
        
        Then we define a credible region of the estimated distribution using the highest density regions (HDR), i.e., the smallest region that
        contains at least $100(1 - \alpha)\%$ of the mass of the inferred photo-z distribution, therefore setting an interval for a given credibility level $(1 - \alpha)$. The expected coverage is the frequency with which the true parameter (\specz) value falls within this highest density region (i.e, inside the calculated interval).
        
        If the \photoz distributions are well calibrated, then the \specz value should be contained in the $1 - \alpha$ highest density regions of the estimated distribution exactly $(1 - \alpha) \times 100\%$ of the time. If the coverage probability is smaller than the credibility level $1 - \alpha$, then this indicates that the $1 - \alpha$ highest density regions are smaller than they should be, which is a sign of overconfident and usually unreliable approximations since it excludes possible physical values for the redshift. On the other hand, if the coverage probability is larger than the credibility level $1 - \alpha$, then this indicates that the highest-density regions are wider than they should be. In this case, the estimated PDFs are said to be conservative.
        
        \begin{figure}
            \centering
            \includegraphics[width=.7\linewidth]{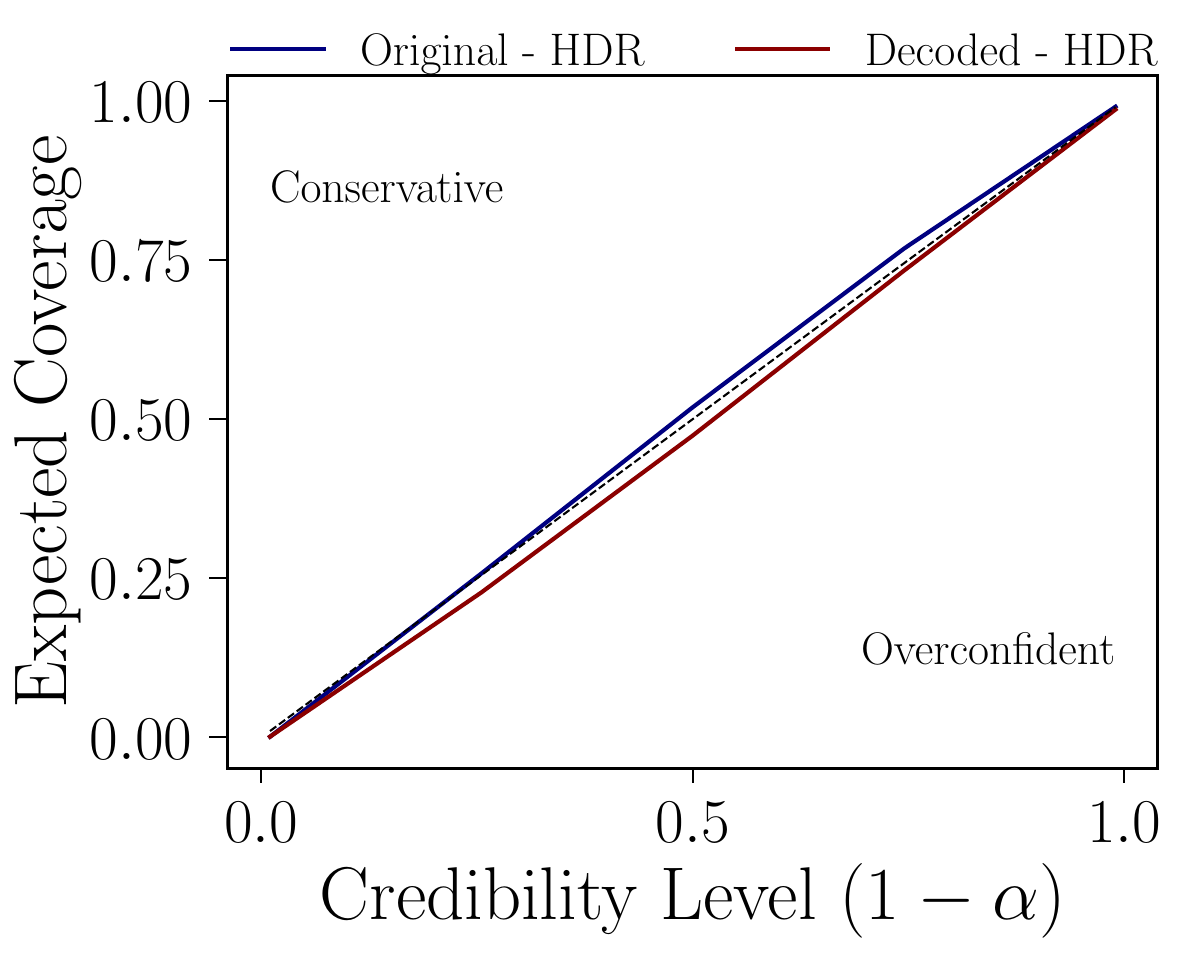}
            \caption{Coverage Test plot: The expected coverage is close to the credibility level $1 - \alpha$, which indicates that the PDFs produced by MDN are neither significantly overdispersed (the coverage curve would otherwise be above the diagonal) nor significantly underdispersed (the coverage curve would be below the diagonal).}
            \label{fig:covertest}
        \end{figure}

        Figure \ref{fig:covertest} shows us that a perfectly calibrated estimated distribution has an expected coverage probability equal to the credibility level. Plotting this relation produces a diagonal line (dashed black line). Conservative models, on the other hand, produce curves above the diagonal and overconfident models underneath.
        The coverage curves (red for the original PDFs from MDN, and navy for the reconstructed PDFs from AE) closely fit the diagonal, which indicates that the PDFs produced by our method are well calibrated \citep{hermans2022trust}, demonstrating the reliability of our method.

    \review{
    \subsection{Efficiency in compressing}
    \label{sec:compressing_results}
        Our results demonstrate that the decoded PDFs from our AE approach can reliably recover the point-like and PDF features from the photometric redshifts' original estimates. To analyze the benefits of using such a technique, we must analyze the efficiency of each approach. 
        
        Figure \ref{fig:time_space_compact} compares two methodologies: generating the PDFs from the AE approach and generating them directly from MDN. This result suggests that time and space in disk consumption for each approach grows following a power law -- in the case of time consumption analysis, this behavior starts from $n_{objects}=10^4$. We fitted a power law function $f(n) = \beta n^{\alpha}$ for all the cases. The coefficients are displaced in Table \ref{tab:power_law_coefficients}. It is worth mentioning that the fit for time consumption considered only the points with $n_{objects}\geq10^4$. 
        
        \begin{figure}
            \centering
            \includegraphics[width=.7\linewidth]{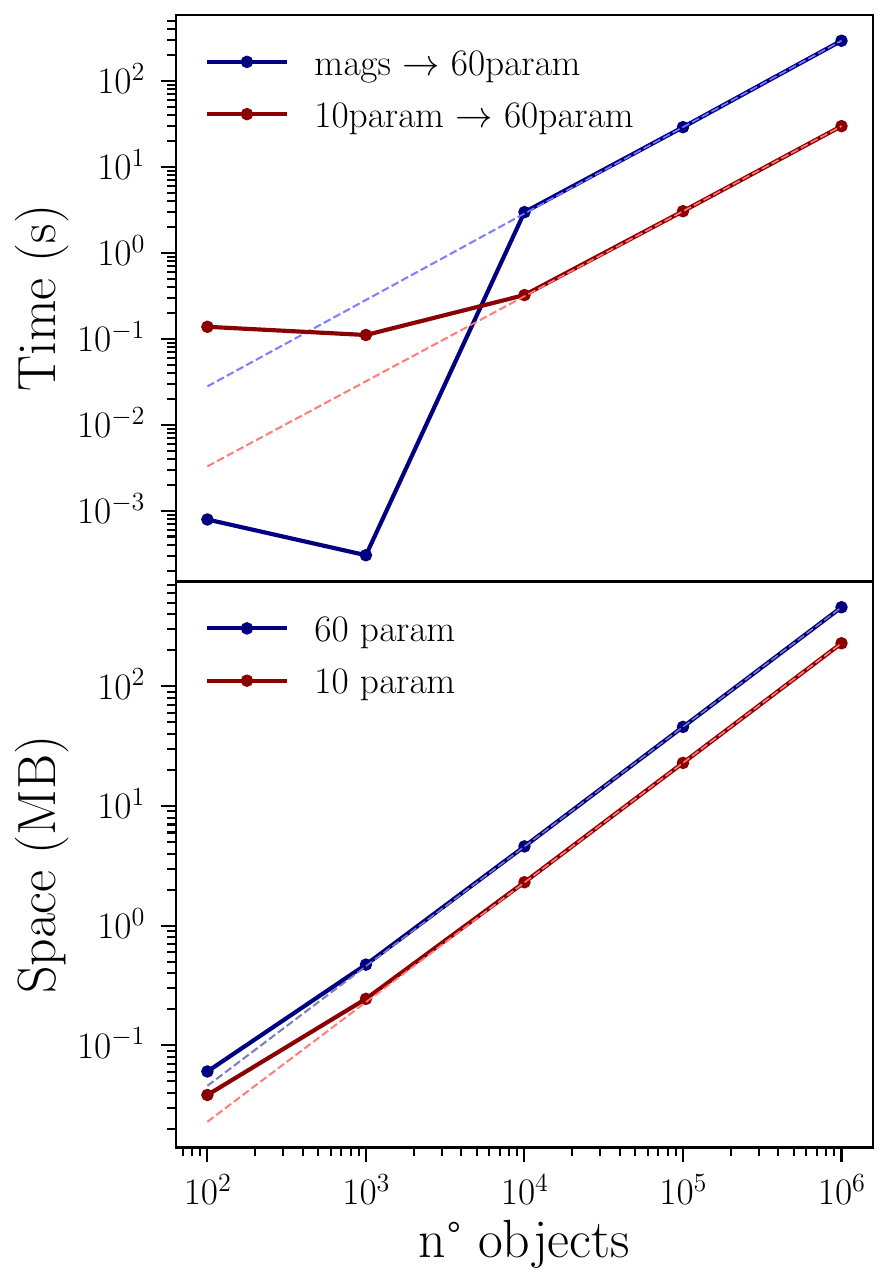}
            \caption{
            \review{The evolution of time (top panel) and space (bottom panel) required to generate and store the PDFs as a function of the number of objects. The relation for the Autoencoder approach is shown in red and the MDN approach in blue. The dashed lines represents the fit of a power law function respective to each process. In both time and space the Autoencoder approach scales better than just the MDN alone.}}

            \label{fig:time_space_compact}
        \end{figure}


    \begin{table}
    \centering{}
    \resizebox{.4\textwidth}{!}{%
    \begin{tabular}{|>{\centering\arraybackslash}m{.3cm}|c|c|c|}
    
      \hline
    
       &  & \textbf{Time} & \textbf{Space}\\
      \hline
      \multirow{2}{*}{\rotatebox[origin=c]{90}{\textbf{MDN}}} & \rule{0pt}{2.2ex} $\alpha$ & $1.004 \pm 0.002$ & $0.9998 \pm 0.0001$ \\ \cline{2-4}
        & \rule{0pt}{2.2ex}  $\beta$ & $(27.2 \pm 0.6)\times 10^{-5}$s  & $(45.87 \pm 0.08)\times 10^{-5}$ MB \\
      \hline
    
      \multirow{2}{*}{\rotatebox[origin=c]{90}{\textbf{AE}}} & \rule{0pt}{2.2ex} $\alpha$ & $0.989 \pm 0.001$ & $0.9996 \pm 0.0003$\\ \cline{2-4}
      & \rule{0pt}{2.2ex} $\beta$ & $(3.44 \pm 0.07)\times 10^{-5}$s & $(22.99 \pm 0.08)\times 10^{-5}$ MB\\
      \hline
    \end{tabular}
    }
    \caption{\centering{Coefficients of the fitted power law functions.}}
    \label{tab:power_law_coefficients}
    \end{table}

        Considering that the $\alpha$ coefficients are close to each other, \textit{i.e.},  the expensiveness for both methods presents the same scaling behavior, 
        we can roughly approximate the ratio between the time and space in disk consumption for both methods by taking the ratio of their respective $\beta$ coefficients. Therefore, from the coefficients' values in Table \ref{tab:power_law_coefficients}, generating the PDFs directly from the magnitudes and colors using the MDN is $7.9 \times$ slower (i.e., it takes $7.9 \times$ more time) then generating them from the encoded 10-sized vector. It also indicates that storing the 60 parameters from the original PDFs is $2.1 \times$ more space-consuming than storing only the 10-sized encoded vectors.    
        For this analysis, we made use of a single core of \texttt{Intel(R) Xeon(R) Platinum 8260 CPU 2.40GHz} with architecture \texttt{x86\_64} when generating the PDFs and used \texttt{astropy} \citep{astropy} tables to store the encoded vectors and the PDF parameters in \texttt{fits} format. 
        }

        \begin{table*}
            \centering{}
            \begin{tabular}{cccccc}
            \hline Training Dataset & Method & Bands used in training & $median\;bias$ &  $\sigma_{NMAD}$& outlier fraction
            \\\hline\hline
            DLTRAIN-A & RF* & GRIZ & 0.005542 & 0.032085 & 0.051842\\
            DLTRAIN-A & MLP* & GRIZ & -0.024966 & 	0.043531 & 	0.055097\\
            DLTRAIN-A & BPZ* & GRIZ & 0.011827 & 0.054326 & 0.104362\\
            DLTRAIN-A & MDN* & GRIZ & -0.001167 & 0.029849 & 0.051821 \\
            -- & Legacy-DR10 method* & -- & 0.0019031& 0.018953 & 0.112017\\

            \textbf{DLTRAIN-A} & \textbf{MDN} & \textbf{GRIZ} & \textbf{-0.001350} & \textbf{0.029345} & \textbf{0.051814}\\
            DLTRAIN-A & MDN & GRI &  -0.003319 & 0.034977 &  0.052609\\   
            DLTRAIN-A & MDN & GRZ & -0.001547 & 0.033075 &  0.058502\\ 
            
            DLTRAIN-B & MDN & GRIZ & -0.005715& 0.029555& 0.048882\\   
            DLTRAIN-C & MDN & GRIZ & -0.002900&  0.028830&  0.049274\\   
            
            DLTRAIN-A & AE & -- & -0.0013288 &0.033287 &  0.050485\\
            \hline
            \end{tabular}
            \caption{\centering{Summary of the metrics computed for all test sample.}}
            \label{tab:metric_results}
        \end{table*}

\section{Discussion and concluding remarks}
\label{sec:discussion}
    \subsection{Summary}
    In this contribution, we introduced a combination between RNNs and MDNs to estimate photometric redshifts \review{in large scale ($>17,000$deg$^2$ of the sky)}. \review{Our approach arises as an innovative way to generate reliable estimates of photometric redshifts using only 3-band coverage.}
    We presented that the quality of our point estimates $z_{phot}$ for DELVE overcomes usual ML techniques and template fitting applied to the same data.
    Our model provides well-calibrated PDFs, which allow error estimations and statistical assessments for any DELVE galaxy.
    
    Storing the PDFs could be hard due to disk space limitations when using this catalog. In that sense, we presented an AE-based technique that compress the 60-sized PDF's parameter array to a 10-sized array. This compression guarantees the accessibility of all the information possible for the DELVE redshifts.
    The final product of this work is one of the most \review{extensive} catalogs \review{of photometric redshifts} with complete probability density estimations, containing $347\rm M$ new measurements of $z_{phot}$ using DELVE photometry.

    \subsection{Mixture Density Networks for photometric redshifts}

        This paper compares a new approach combining the MDN technique and LMU cells to other ML methods for photometric redshift estimations on DELVE-DR2 data. The main reason for using MDNs is the capability to generate mathematically well-defined PDFs for photo-z. \review{This promotes} possible analytical procedures, \review{facilitates estimating uncertainties} and can be validated by several \review{diagnostics}.
        In Section \ref{subsec:point_results}, we explored the influence of the $z$ distribution in the training set on the final results. We show that even using less data than the other training sets, \review{the best results came from the A-trained model (DLTRAIN-A)}, showing that the homogeneity of the training set is \review{critical} to avoid significant biases on the high-z region, \review{agrreing with the results from} \citet{Zou_2019}. 
        
        Furthermore, the MDN proved a potentially good estimator for objects with \review{detections in only 3 photometric bands}. Figure \ref{fig:coverage_metrics} shows a divergence in outlier fraction for the model trained with \textit{g,r,z} bands around $z_{spec}\approx0.3$. However, considering the error bars, the results are entirely compatible with those derived using 4-band coverage. It is worth remembering that the amount of 3-band data with spectroscopic redshift was insufficient to produce or improve any training set for the model. Therefore, we must be aware that the model learned how to estimate redshifts using 3 bands from a set of galaxies with 4 band detections. In that sense, any physical information intrinsic to having a missing band can not be interpreted for the model when applied to the DELVE-DR2 catalog.

       In our comparative analysis, we evaluated the performance of several techniques for $z_{phot}$ estimation. The MDN outperformed other methods across multiple metrics (see Table \ref{tab:metric_results}). For instance, when averaging the entire test dataset, our model presented the lowest absolute median bias, the lowest scatter, and the second lowest outlier fraction relative to the ML and template fitting models. 
        For the metrics derived from the public LS-DR10 photometric redshifts, the absolute median bias and scatter, respectively, achieved values 12\% and 38\% lower than the ones derived from MDN. Nonetheless, the presence of outliers is 108\% higher in the public available redshifts relative to the MDN ones. 

        In section \ref{subsec:pdf_anal} 
        we evaluate the quality of the estimated redshift distribution using three diagnostic tools: The Probability Integral Transform, the Odds distributions, and the Coverage Diagnostic.
        Our model did not demonstrate any strong systematic underestimation or overestimation of the PDF's likeliest values. Additionally,  the Odds and coverage diagnostics evidence that our model produces calibrated redshift distribution, i.e., it did not demonstrate a systematic overestimation or underestimation of the PDF's variance. Therefore, our method generates reliable PDFs for scientific use.

        We show that, in addition to generating well-calibrated PDFs, the point estimation of \photoz outperformed other well-known algorithms. \review{And these estimates are} comparable to the public redshifts available from the latest data release of \review{DECaLS} Survey. 
        \review{This comparable performance is achieved even though DECALs employs a different survey strategy, photometric calibration, and uses 5 bands \citep{Zou_2019}.}
        
        


    \subsection{Probability Density Functions availability}

        The DR2 is the latest public release from DELVE. 
        However, as successive data releases encompass progressively deeper regions in  the sky, \review{there will be a drastic increase in the number of observed galaxies}.
        Beyond DELVE, the cumulative impact of advancements in optical surveys imaging drives astronomy to the domain of big data. In the context of $z_{phot}$'s PDFs, storing their complete information for hundreds of thousands of galaxies can be challenging, depending on the model used to generate them.
        
        In this work, we present a novel approach for compressing the PDF's information derived from MDN using an \review{AE}. This methodology provides two main contributions: firstly, it significantly reduces the space in disk occupied for each PDF by compressing the 60-parameter array defining the PDF to a 10-params array that represents the PDF information in the latent space of the AE; the second contribution lies in the fact that the AE model is less complex than the MDN, i.e., it is faster to generate the PDFs from the 10-params vector then run the MDN directly from the magnitudes and colors. \review{We show in Section \ref{sec:compressing_results} that using the AE approach lead to a PDF generation $7.9$ times faster and twice less space-consuming.}

        In summary, the AE approach demonstrates the capability to reconstruct the PDFs generated from MDN while preserving their quality. This strategy facilitates greater accessibility \review{of} the entire information from the PDFs, requiring less demanding hardware settings.
        It is worth noting that the technique is not perfect, as shown in Section \ref{sec:results}. Nonetheless, it reproduces accurately the point estimates despite minor deviations from the PIT and Odds distributions. These deviations may be addressed by investigating the calibration of the AE in future works.
        Future efforts will also focus on enhancing this approach by applying Auto-Machine Learning techniques and exploring different DL architectures and pre-processing methods.
        


    \subsection{A new photometric redshift catalogue with full PDFs}
    \label{sec:new_catatlogue}
    
    \citet{Lima_2022} provides complete PDFs using a similar MDN approach for the S-PLUS first data release, which covers around $336\text{deg}^2$ of the Stripe 82 region. The second data release of S-PLUS features improved photometry and $z_{phot}$'s PDFs for around $950.5\text{deg}^2$ of the sky \citep{almeida2022}. \citet{Duncan_2022} produces good quality redshifts for the LS-DR8, one of the biggest footprints of sky imaging publicly available. They use GPz to estimate $z_{phot}$ for $\approx 9.5 \times 10^8$ galaxies within $19 400\;deg^2$ (both hemispheres) in the sky, with redshifts values spanning between $0<z<6$.   
    
    The final products of this work consist of: new $z_{phot}$'s PDFs for $347 M$ galaxies within the DELVE-DR2 footprint; a novel approach combining LMU architecture with Mixture Density Models capable of estimating new $z_{phot}$ PDFs for upcoming data from DELVE; an AE-based method to reduce the disk space occupied by the PDFs, allowing for their accurate reconstruction, and thus enabling an accessible way to characterize $z_{phot}$ statistics in terms of PDF. The photometric redshifts produced by our approaches present a marginal increase in bias at some values of $z$ when compared to the LS-DR10 ones, for example (Section \ref{sec:discussion}). However, we also provide a way to access the complete probabilistic information about each galaxy in DELVE-DR2.  From this point of view, this contribution provides one of the vastest existing photometric redshift catalogs \review{for} the \review{southern} hemisphere paired with full probability density estimates.

    The utilization of our photometric redshift catalog by \citet{alfradique2024} in their \review{Dark} Sirens analysis, \review{i.e., analysis on the gravitational signal from binary black hole mergers,} highlights its significance. 
    \review{They used our catalog to constrain the Universe's expansion rate, yielding a value of $H_{0} = 68.00^{+4.28}_{-3.85} \rm {km s^{-1} Mpc^{-1}}$. Crucially, this analysis relied on the complete PDF information for galaxies, not just a point estimate. This underscores the critical importance of the complete PDF information for galaxies. Since PDFs may exhibit multi-modal distributions rather than conform to normal distributions, their nuanced nature can enrich subsequent analyses, facilitating deeper insights into cosmic phenomena.}
    

    
    \review{ The final catalog features photometric redshift values as a key enhancement. These $z_{phot}$ values hold significant potential for various astrophysical research applications, including gravitational wave follow-ups, dark siren cosmology, large-scale structure analysis, and environmental studies. }
    


    
    \subsection{Future challenges for Deep Learning photometric redshifts}
        
        Deep learning has emerged as a promising avenue for estimating photometric redshifts due to its ability to learn complex patterns in large datasets. However, there are inherent challenges in building reliable DL models that encapsulate photometric redshifts' full probability density functions (PDFs). These challenges include:

        \review{
        \begin{itemize}
             \item Addressing biases and systematics in the data, which can adversely affect the performance of the models.

            \item Quantifying uncertainties in the photometric redshift estimates, which is essential for understanding their reliability and propagation of errors in subsequent analyses.

            \item Dealing with catastrophic outliers, which can significantly impact the accuracy and precision of photometric redshifts.

            \item Ensuring the models' interpretability, which is vital for understanding the underlying astrophysical processes.
        \end{itemize}
        } 
        
    In conclusion, photometric redshifts have been and will continue to be indispensable in wide-field surveys, such as DES, DECaLS, DELVE, and the upcoming Vera C. Rubin LSST. Developing reliable DL models that encapsulate the full PDFs of photometric redshifts remains an ongoing challenge, with potential solutions involving combining state-of-the-art machine learning techniques and rigorous astrophysical understanding. Overcoming these challenges will pave the way for more accurate and reliable photometric redshift estimates, ultimately enabling a deeper understanding of the Universe's structure and evolution.
    The increased depth and coverage of LSST will enable the refinement of photometric redshift techniques, opening up new avenues for cosmological research and potentially transforming our understanding of the Universe.

\section{Data Availability}
\label{data_acess}
\review{
All the \photoz data described in \ref{sec:catalog} will be made publicly available soon in the Astro Data Lab DELVE DR2 web page \footnote{\hyperlink{https://datalab.noirlab.edu/delve}{https://datalab.noirlab.edu/delve}}. We are also managing to make both the encoded PDFs and the decoder model publicly available. Updates about this will be posted in the GitGub profile \hyperlink{https://github.com/gsmteixeira}{gsmteixeira}.}

\section*{Acknowledgements}
C. Bom acknowledges the financial support from CNPq (316072/2021-4) and from FAPERJ (grants 201.456/2022 and 210.330/2022) and the FINEP contract 01.22.0505.00 (ref. 1891/22). The authors made use of Sci-Mind servers machines developed by the CBPF AI LAB team and would like to thank A. Santos, P. Russano, and M. Portes de Albuquerque for all the support in infrastructure matters. 

G. Teixeira aknowleges the financial support from CNPq (PhD fellowship, 140212/2022-1) and from FAPERJ (PhD merit fellowship - FAPERJ NOTA 10, 202.432/2024).

The DELVE project is partially supported by Fermilab LDRD project L2019-011 and the NASA Fermi Guest Investigator Program Cycle 9 No. 91201. This project used data obtained with the Dark Energy Camera (DECam), which was constructed by the Dark Energy Survey (DES) collaboration. Funding for the DES Projects has been provided by the US Department of Energy, the U.S. National Science Foundation, the Ministry of Science and Education of Spain, the Science and Technology Facilities Council of the United Kingdom, the Higher Education Funding Council for England, the National Center for Supercomputing Applications at the University of Illinois at Urbana-Champaign, the Kavli Institute for Cosmological Physics at the University of Chicago, Center for Cosmology and Astro-Particle Physics at the Ohio State University, the Mitchell Institute for Fundamental Physics and Astronomy at Texas A\&M University, Financiadora de Estudos e Projetos, Fundação Carlos Chagas Filho de Amparo à Pesquisa do Estado do Rio de Janeiro, Conselho Nacional de Desenvolvimento Científico e Tecnológico and the Ministério da Ciência, Tecnologia e Inovação, the Deutsche Forschungsgemeinschaft and the Collaborating Institutions in the Dark Energy Survey.
The Collaborating Institutions are Argonne National Laboratory, the University of California at Santa Cruz, the University of Cambridge, Centro de Investigaciones Enérgeticas, Medioambientales y Tecnológicas–Madrid, the University of Chicago, University College London, the DES-Brazil Consortium, the University of Edinburgh, the Eidgenössische Technische Hochschule (ETH) Zürich, Fermi National Accelerator Laboratory, the University of Illinois at Urbana-Champaign, the Institut de Ciències de l’Espai (IEEC/CSIC), the Institut de Física d’Altes Energies, Lawrence Berkeley National Laboratory, the Ludwig-Maximilians Universität München and the associated Excellence Cluster Universe, the University of Michigan, NSF NOIRLab, the University of Nottingham, the Ohio State University, the OzDES Membership Consortium, the University of Pennsylvania, the University of Portsmouth, SLAC National Accelerator Laboratory, Stanford University, the University of Sussex, and Texas A\&M University.
This project is based on observations at NSF Cerro Tololo Inter-American Observatory, NSF NOIRLab (NOIRLab Prop. ID 2019A-0305; PI: Alex Drlica-Wagner), which is managed by the Association of Universities for Research in Astronomy (AURA) under a cooperative agreement with the U.S. National Science Foundation.
This manuscript has been authored by Fermi Research Alliance, LLC, under contract No. DE-AC02-07CH11359 with the US Department of Energy, Office of Science, Office of High Energy Physics. The United States Government retains and the publisher, by accepting the article for publication, acknowledges that the United States Government retains a non-exclusive, paid-up, irrevocable, worldwide license to publish or reproduce the published form of this manuscript, or allow others to do so, for United States Government purposes
\notes{INCLUDING LEGACY DR10 ACKNOWLEDGEMENTS}

The Legacy Surveys consist of three individual and complementary projects: the Dark Energy Camera Legacy Survey (DECaLS; Proposal ID \#2014B-0404; PIs: David Schlegel and Arjun Dey), the Beijing-Arizona Sky Survey (BASS; NOAO Prop. ID \#2015A-0801; PIs: Zhou Xu and Xiaohui Fan), and the Mayall z-band Legacy Survey (MzLS; Prop. ID \#2016A-0453; PI: Arjun Dey). DECaLS, BASS and MzLS together include data obtained, respectively, at the Blanco telescope, Cerro Tololo Inter-American Observatory, NSF’s NOIRLab; the Bok telescope, Steward Observatory, University of Arizona; and the Mayall telescope, Kitt Peak National Observatory, NOIRLab. Pipeline processing and analyses of the data were supported by NOIRLab and the Lawrence Berkeley National Laboratory (LBNL). The Legacy Surveys project is honored to be permitted to conduct astronomical research on Iolkam Du’ag (Kitt Peak), a mountain with particular significance to the Tohono O’odham Nation.

NOIRLab is operated by the Association of Universities for Research in Astronomy (AURA) under a cooperative agreement with the National Science Foundation. LBNL is managed by the Regents of the University of California under contract to the U.S. Department of Energy.

This project used data obtained with the Dark Energy Camera (DECam), which was constructed by the Dark Energy Survey (DES) collaboration. Funding for the DES Projects has been provided by the U.S. Department of Energy, the U.S. National Science Foundation, the Ministry of Science and Education of Spain, the Science and Technology Facilities Council of the United Kingdom, the Higher Education Funding Council for England, the National Center for Supercomputing Applications at the University of Illinois at Urbana-Champaign, the Kavli Institute of Cosmological Physics at the University of Chicago, Center for Cosmology and Astro-Particle Physics at the Ohio State University, the Mitchell Institute for Fundamental Physics and Astronomy at Texas A\&M University, Financiadora de Estudos e Projetos, Fundacao Carlos Chagas Filho de Amparo, Financiadora de Estudos e Projetos, Fundacao Carlos Chagas Filho de Amparo a Pesquisa do Estado do Rio de Janeiro, Conselho Nacional de Desenvolvimento Cientifico e Tecnologico and the Ministerio da Ciencia, Tecnologia e Inovacao, the Deutsche Forschungsgemeinschaft and the Collaborating Institutions in the Dark Energy Survey. The Collaborating Institutions are Argonne National Laboratory, the University of California at Santa Cruz, the University of Cambridge, Centro de Investigaciones Energeticas, Medioambientales y Tecnologicas-Madrid, the University of Chicago, University College London, the DES-Brazil Consortium, the University of Edinburgh, the Eidgenossische Technische Hochschule (ETH) Zurich, Fermi National Accelerator Laboratory, the University of Illinois at Urbana-Champaign, the Institut de Ciencies de l’Espai (IEEC/CSIC), the Institut de Fisica d’Altes Energies, Lawrence Berkeley National Laboratory, the Ludwig Maximilians Universitat Munchen and the associated Excellence Cluster Universe, the University of Michigan, NSF’s NOIRLab, the University of Nottingham, the Ohio State University, the University of Pennsylvania, the University of Portsmouth, SLAC National Accelerator Laboratory, Stanford University, the University of Sussex, and Texas A\&M University.

BASS is a key project of the Telescope Access Program (TAP), which has been funded by the National Astronomical Observatories of China, the Chinese Academy of Sciences (the Strategic Priority Research Program “The Emergence of Cosmological Structures” Grant \# XDB09000000), and the Special Fund for Astronomy from the Ministry of Finance. The BASS is also supported by the External Cooperation Program of Chinese Academy of Sciences (Grant \# 114A11KYSB20160057), and Chinese National Natural Science Foundation (Grant \# 12120101003, \# 11433005).

The Legacy Survey team makes use of data products from the Near-Earth Object Wide-field Infrared Survey Explorer (NEOWISE), which is a project of the Jet Propulsion Laboratory/California Institute of Technology. NEOWISE is funded by the National Aeronautics and Space Administration.

The Legacy Surveys imaging of the DESI footprint is supported by the Director, Office of Science, Office of High Energy Physics of the U.S. Department of Energy under Contract No. DE-AC02-05CH1123, by the National Energy Research Scientific Computing Center, a DOE Office of Science User Facility under the same contract; and by the U.S. National Science Foundation, Division of Astronomical Sciences under Contract No. AST-0950945 to NOAO.

\appendix
\section{Catalogue Description}
\label{sec:catalog}
The photo-z catalog compiles a comprehensive set of redshift measurements using the photometric bands available in the DECam Local Volume Exploration survey. The QUICK\_OBJECT\_ID uniquely identifies each entry in the catalog, followed by the corresponding celestial coordinates (RA and DEC).

The primary redshift estimation comes from the Z\_PHOT\_PEAK. This variable represents the peak of the probability density function (PDF) and can be used as the main photo-z value. The photo-z uncertainty is given by Z\_PHOT\_ERR, representing the interval encompassing 68$\%$ (1 sigma) of the PDF, commonly known as sigma 68. Additionally, Z\_PHOT\_MEDIAN provides the PDF's median, offering an alternative for extracting the redshift value.

The Z\_PHOT\_ODDS values, as described in section \ref{sec:results}, offer insights into the reliability of the redshift estimation. Higher Z\_PHOT\_ODDS values indicate more robust redshift predictions. The catalog also includes the Z\_PHOT\_SAMPLES, featuring three random values selected from the PDF. It can give the user an idea of the PDF distribution. 

The percentiles Z\_PHOT\_L68 and Z\_PHOT\_U68 represent the 16th and 84th percentiles of the PDF, respectively, offering a range covering most of the probable redshift values. These percentiles contribute to understanding the uncertainties associated with the redshift estimations. The boolean variables  'MODEL\_GRIZ,' 'MODEL\_GRI,' and 'MODEL\_GRZ' identify whether the redshift was predicted using models that use combinations of 4 or 3 bands. These models contribute to the richness of the catalog, allowing users to access photo-z measurements even if the target was not fully observed in the 4 photometric bands.

In summary, our photometric redshift catalog provides information related to central estimates,  uncertainties, reliability metrics, and model-specific predictions, which the astronomical community can use for different scientific cases using DELVE data. A general overview of the columns available in the catalog is presented in Table  \ref{tab:cat_description_long}.

\begin{table*}[htbp]
\caption{DELVE DR2 photometric redshift tables: column descriptions}
\centering
\begin{tabular}{p{0.20\linewidth} p{0.6\linewidth} p{0.15\linewidth}}
    \hline
    \textbf{Column Name} & \textbf{Description} & \textbf{Columns} \\
    \hline
    QUICK\_OBJECT\_ID & Unique identifier for each object & 1 \\
    RA & Right ascension from DELVE catalogs (degrees) & 1 \\
    DEC & Declination from DELVE catalogs (degrees) & 1 \\
    Z\_PHOT\_PEAK & Peak value of the PDF, considered as the true photo-z value & 1 \\
    Z\_PHOT\_ERR & Photo-z uncertainty represented by the interval that encompasses 68\% of the PDF, denoted as sigma or the 1$\sigma$ percentile & 1 \\
    Z\_PHOT\_MEDIAN & Photo-z measurement based on the median PDF value & 1 \\
    Z\_PHOT\_ODDS & Odds values as described in Section \ref{sec:results} & 1 \\
    Z\_PHOT\_SAMPLES & Three photo-z values randomly selected from the PDF & 1 \\
    Z\_PHOT\_L68 & 16\% percentile & 1 \\
    Z\_PHOT\_U68 & 84\% percentile & 1 \\
    MODEL\_GRIZ & Boolean: True if predicted by the GRIZ model (i.e., using four photometric bands) & 1 \\
    MODEL\_GRI & Boolean: True if predicted by the GRI model (i.e., using only three photometric bands) & 1 \\
    MODEL\_GRZ & Boolean: True if predicted by the GRZ model (i.e., using only three photometric bands) & 1 \\
    \hline
\end{tabular}
\label{tab:cat_description_long}
\end{table*}

\end{document}